\documentclass[pdflatex,sn-mathphys-num]{sn-jnl}

\usepackage{rotating} 
\usepackage{booktabs} 
\usepackage{graphicx}%
\usepackage{multirow}%
\usepackage{amsmath,amssymb,amsfonts}%
\usepackage{amsthm}%
\usepackage{mathrsfs}%
\usepackage{siunitx}
\usepackage[title]{appendix}%
\usepackage{xcolor}%
\usepackage{textcomp}%
\usepackage{manyfoot}%
\usepackage{booktabs}%
\usepackage{algorithm}%
\usepackage{algorithmicx}%
\usepackage{algpseudocode}%
\usepackage{listings}%
\usepackage{bm}

\theoremstyle{thmstyleone}%
\theoremstyle{thmstyletwo}%

\theoremstyle{thmstylethree}%

\begin{document}

\title[PICS for coupled PDEs]{PICS: A Partition-of-unity Information-geometric Certified Solver for Coupled Partial Differential Equations}


\author[1]{\fnm{Ze}\sur{Tao}}
\author[1]{\fnm{Hongfu}\sur{Zhou}}
\author[1]{\fnm{Hanbing} \sur{Liang}}
\author*[1]{\fnm{Fujun}\sur{Liu}}\email{fjliu@cust.edu.cn}

\affil[1]{Nanophotonics and Biophotonics Key Laboratory of Jilin Province, School of Physics, Changchun University of Science and
Technology, Changchun, 130022, P.R. China}


\abstract{Coupled partial differential equations underpin a wide range of multiphysics systems, yet existing neural PDE solvers still struggle to resolve localized high-risk regions and often fail to preserve structural admissibility across coupled fields. To address these limitations, we propose the Partition-of-unity Information-geometric Certified Solver (PICS), a closed-loop framework that strictly enforces structural admissibility at the level of representation rather than relying on an additional soft penalty. By constructing a gate-structured admissible manifold coupled with a restricted jet prolongation, PICS ensures that geometry-sensitive approximations and closure-essential differential coordinates enter the solver as a strongly enforced, structure-preserving ansatz. Furthermore, the framework integrates entropic tail-risk control and \textit{a posteriori} certificate-driven empirical measure transport, dynamically reallocating training efforts toward uncertified, error-prone transition zones. Evaluated against standard baseline methods across three two-dimensional coupled benchmarks, PICS achieves more consistently accurate and balanced cross-field recovery while retaining practical computational efficiency, thereby providing a rigorous route toward highly reliable multiphysics simulation.}




\maketitle

\section*{Introduction}

Coupled partial differential equations govern a broad class of transport, electro-thermal, and multiphysics processes\cite{karniadakis2021piml,cuomo2022sciml} in which admissibility constraints, localized transition layers, and cross-field scale imbalance act simultaneously. In such settings, a practically useful solver must recover multiple fields in a mutually consistent manner, preserve structural constraints at the representation level\cite{jagtap2020xpinns,moseley2023fbpinns}, and suppress spatially concentrated error\cite{nabian2021importance,wu2023sampling} rather than merely reduce an averaged discrepancy. These requirements become particularly stringent when the solution exhibits geometry-active regions, interface-type transitions, or strongly uneven residual distributions\cite{sarma2024ipinns,jagtap2020cpinn}. A solver that performs well only in a mean sense but loses control of high-risk regions\cite{anagnostopoulos2024rba,mcclenny2023sapinn} remains inadequate for robust multiphysics recovery, especially when strict conservation laws must be upheld.

Recent neural PDE solvers\cite{lu2021deeponet,kovachki2023neuraloperator,wang2021pideeponet} have demonstrated considerable flexibility for mesh-free approximation and physics-guided optimization, including representative residual-driven frameworks such as physics-informed neural networks (PINN)\cite{raissi2019pinn}, the Deep Galerkin Method (DGM)\cite{sirignano2018dgm}, and the Deep Ritz Method (DRM)\cite{yu2018drm}. Nevertheless, three difficulties remain persistent. First, a single global approximation class often lacks sufficient geometric sensitivity to represent localized transition structures and heterogeneous active regions in a balanced way\cite{jagtap2020adaptiveactivation,dolean2024multilevel}. Second, standard mean-type residual objectives typically treat physical constraints as soft penalties, leading to non-physical mass leakage and failing to directly control worst-point behavior, so a visually acceptable global field may still contain concentrated local failure\cite{krishnapriyan2021failure,wang2022ntk}. Third, static or weakly adaptive collocation policies do not continuously reallocate training effort toward the evolving high-risk regions revealed by the solver itself\cite{wang2024causality,mao2023adaptive}. These three deficiencies jointly limit cross-field consistency, tail-error suppression, and robustness under case-dependent closure changes\cite{wang2021gradientpathologies}.

To address these issues, we introduce the \emph{Partition-of-unity Information-geometric Certified Solver} (PICS), a gate-structured admissible-manifold solver equipped with restricted jet prolongation, entropic residual geometry, and certificate-driven empirical measure transport. At the representation level, PICS constructs a gate-structured admissible manifold that couples local chart families through a fixed partition mechanism, thereby providing a geometry-aware approximation class that enforces critical physical conditions as hard constraints. By leveraging a streamfunction-based latent representation, exact mass conservation is built natively into the admissible state, bypassing notorious penalty parameter tuning and providing a strongly enforced, structure-preserving ansatz. At the closure level, PICS retains only the finite derivative coordinates required by the governing system through restricted jet prolongation, ensuring that the closure structure itself becomes an explicit algorithmic object. Furthermore, at the objective and sampling levels, PICS integrates mean residual minimization, entropic tail-risk control, and boundary or interface consistency. Crucially, this residual stage evaluates a normalized section to extract a spatial certificate field, which acts as an \textit{a posteriori} error estimator. This estimator then drives an empirical measure transport mechanism, continuously reallocating training samples toward uncertified extreme regions in a manner conceptually analogous to adaptive mesh refinement (AMR) in traditional computational mechanics.

We formulate PICS as a closed solver framework acting on admissible states, residual sections, certificate fields, and transported empirical measures, evaluating it against PINN, DGM, and DRM across three coupled benchmark PDE systems. These benchmarks encompass challenging closure modifications, including thermo-viscous Leray-regularized transport that conceptually relates to subgrid-scale turbulence modeling, and pressure-regularized screened electro-thermal coupling frequently encountered in singular perturbation limits. The main text presents the solver principle, the aggregate quantitative comparison, and the three representative cases in concise form, while the Supplementary Information provides the full definitions, propositions, proofs, and parameter analyses for the admissible manifold, restricted jet prolongation, entropic residual geometry, certificate field, persistent risk memory, and empirical measure transport.
\section*{Results}

\subsection*{PICS principle and solver overview}

\begin{figure*}[!ht]
	\centering
	\includegraphics[width=1.15\textwidth]{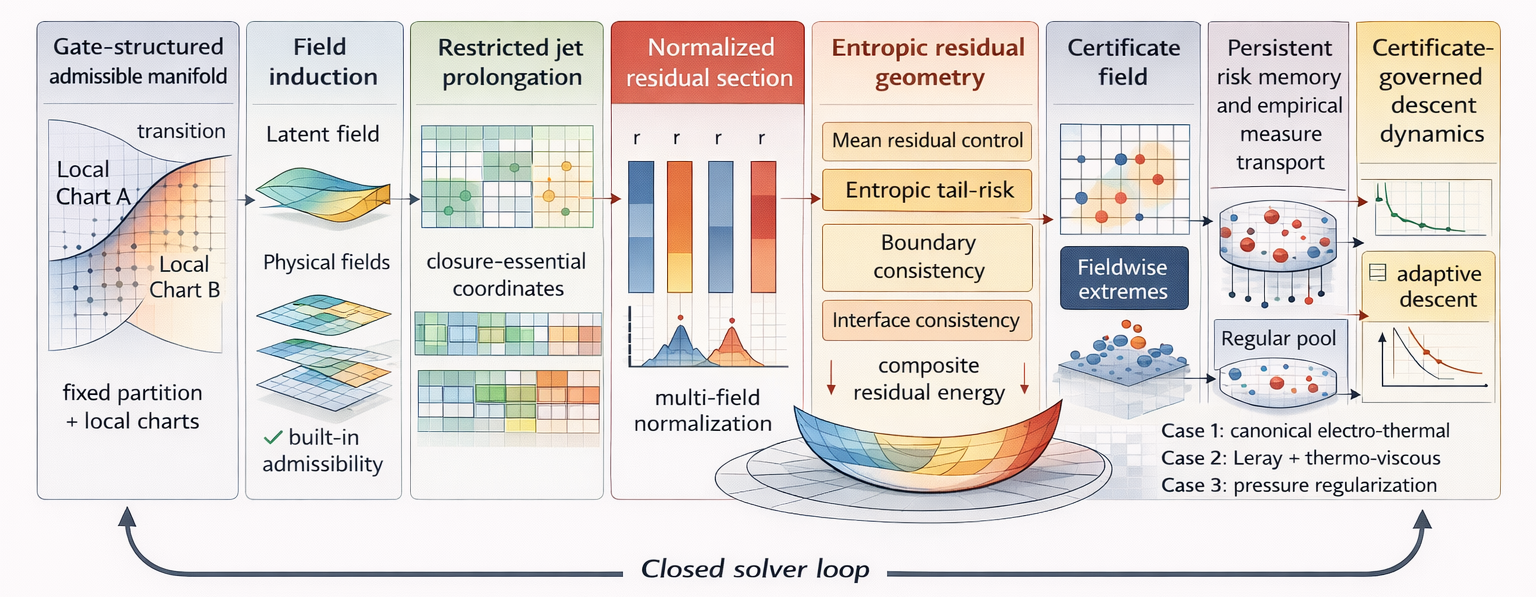}
\caption{\textbf{The framework of the proposed PICS solver.} PICS organizes PDE solving as a closed solver loop rather than a residual-fitting pipeline. Starting from a gate-structured admissible state, the solver induces the case-level physical fields, retains only the closure-essential differential coordinates through a restricted jet prolongation, evaluates a normalized residual section and its composite residual geometry, extracts a certificate field that identifies fieldwise and global high-risk regions, transports the empirical training measure toward those regions, and then updates the parameter state on the transported measure. This common loop serves as the solver backbone for all benchmark systems studied in Figs.~\ref{F2}--\ref{F4} and Tables~\ref{T1}--\ref{T2}.}
	\label{F1}
\end{figure*}

Fig.~\ref{F1} presents PICS as a closed solver loop rather than a residual-fitting pipeline. Starting from a gate-structured admissible state, the solver induces the physical fields, retains only closure-essential differential coordinates through a restricted jet prolongation, evaluates a normalized residual section, builds a composite residual geometry, extracts a certificate field, transports the empirical training measure, and updates the parameter state. This loop serves as the common solver backbone for all benchmark systems reported in Figs.~\ref{F2}--\ref{F4} and Tables~\ref{T1}--\ref{T2}. For each case \(c\), we write the benchmark system on \(\Omega_c\) with boundary operator on \(\partial\Omega_c\) as
\begin{equation}
\mathcal{F}_c[q]=0
\quad \text{in } \Omega_c,
\qquad
\mathcal{G}_c[q]=0
\quad \text{on } \partial\Omega_c,
\label{eq:pics-benchmark-family}
\end{equation}
where \(q\) denotes the case-level solver state. When a case contains an internal interface, we denote it by \(\Gamma_c\subset \overline{\Omega}_c\); otherwise we set \(\Gamma_c=\varnothing\). The solver objects introduced below are attached to Eq.~\eqref{eq:pics-benchmark-family}.

PICS restricts the solver state to a gate-structured admissible manifold,
\begin{equation}
\mathcal{M}_c
=
\left\{
q_\theta
=
(1-g_c)\,q_{L,\theta}
+
g_c\,q_{R,\theta}
\;:\;
\theta\in\Theta_c
\right\},
\label{eq:pics-admissible-manifold}
\end{equation}
where \(q_\theta\) denotes the gate-structured admissible state, \(g_c\) denotes a fixed partition weight associated with case \(c\), and \(q_{L,\theta}\), \(q_{R,\theta}\) denote local chart families. Eq.~\eqref{eq:pics-admissible-manifold} gives PICS a structured state space before any residual evaluation starts. Detailed geometric and regularity properties of \(\mathcal{M}_c\) are given in Section~II of the Supplementary Materials (SM). The physical fields then enter through the field-induction map \(\mathfrak{P}\). For the benchmark systems considered here, we write the admissible state as
\begin{equation}
q_\theta
=
\bigl(\psi_\theta,\pi_\theta,\varphi_\theta,\tau_\theta\bigr),
\qquad
\mathfrak{P}(q_\theta)
=
\bigl(u_\theta,v_\theta,p_\theta,\phi_\theta,T_\theta\bigr)
=
\bigl(\partial_y\psi_\theta,\,-\partial_x\psi_\theta,\,\pi_\theta,\,\varphi_\theta,\,\tau_\theta\bigr).
\label{eq:pics-field-induction}
\end{equation}
Here, $\psi_{\theta}$ acts as the latent stream component. Instead of operating directly on the physical velocity vector, PICS induces the velocity field inherently from this latent streamfunction. Consequently, the induced velocity field satisfies
\begin{equation}
\partial_{x}u_{\theta}+\partial_{y}v_{\theta}=\partial_{xy}\psi_{\theta}-\partial_{yx}\psi_{\theta}=0,
\label{eq:pics-incompressibility}
\end{equation}
whenever the mixed derivatives commute. PICS therefore strictly enforces structural admissibility at the level of representation. By providing a strongly enforced, divergence-free ansatz, this mechanism fundamentally avoids the extensive hyperparameter tuning typically required for soft-penalty divergence losses in standard neural solvers. Section~III gives the full field-induction construction, and Section~VIII states the corresponding admissibility and consistency propositions. PICS then acts on a restricted jet object rather than an unrestricted differential closure. We define the restricted jet prolongation by
\begin{equation}
\mathfrak{j}_c[q_\theta](x)
=
\bigl(
\partial^\alpha q_\theta(x)
\bigr)_{\alpha\in\mathcal{A}_c},
\label{eq:pics-restricted-jet}
\end{equation}
where \(\mathcal{A}_c\) denotes a finite case-dependent multi-index set. Only closure-essential coordinates are retained in Eq.~\eqref{eq:pics-restricted-jet}; PICS therefore keeps the differential content required by the case-dependent PDE family and discards the remainder. Section~IV of the SM lists the complete case-wise closure sets and the associated restricted jet structure. The residual stage starts from a normalized residual section,
\begin{equation}
\widehat{\mathfrak{R}}_c[q_\theta](x)
=
\widehat{\mathbf r}_{c,\theta}(x)
=
\bigl(
\widehat r_{c,\theta,1}(x),\ldots,\widehat r_{c,\theta,M_c}(x)
\bigr),
\label{eq:pics-normalized-residual-section}
\end{equation}
with channelwise normalization
\begin{equation}
\widehat r_{c,\theta,m}(x)
=
\frac{r_{c,\theta,m}(x)}{S_{c,m}},
\qquad
S_{c,m}>0,
\qquad
m=1,\ldots,M_c,
\label{eq:pics-residual-normalization}
\end{equation}
where \(r_{c,\theta,m}\) denotes the \(m\)-th raw residual channel and \(S_{c,m}\) denotes its case-dependent reference scale. Thus PICS evaluates a normalized residual section rather than a raw residual stack. Section~V of the SM gives the detailed case-wise residual constructions and the normalization rationale. The normalized residual section induces the composite residual energy
\begin{equation}
\mathcal{E}_c(\theta;\mu_k)
=
\mathcal{E}^{\mathrm{mean}}_c(\theta;\mu_k)
+
\lambda_{\mathrm{ent}}\,\mathcal{E}^{\mathrm{tail}}_c(\theta;\mu_k)
+
\lambda_{\partial\Omega}\,\mathcal{B}_c(\theta)
+
\lambda_{\Gamma}\,\mathcal{I}_c(\theta;\mu_k),
\label{eq:pics-composite-energy}
\end{equation}
which matches the four central control blocks in Fig.~\ref{F1}: mean residual control, entropic tail-risk, boundary consistency, and interface consistency. The mean term reads
\begin{equation}
\mathcal{E}^{\mathrm{mean}}_c(\theta;\mu_k)
=
\int_{\Omega_c}
\bigl\|
\widehat{\mathbf r}_{c,\theta}(x)
\bigr\|_2^2
\,\mathrm{d}\mu_k(x),
\label{eq:pics-mean-energy}
\end{equation}
and the entropic tail-risk term reads
\begin{equation}
\mathcal{E}^{\mathrm{tail}}_c(\theta;\mu_k)
=
\frac{1}{\beta_c}
\log
\left(
\int_{\Omega_c}
\exp\!\Bigl(
\beta_c
\bigl\|
\widehat{\mathbf r}_{c,\theta}(x)
\bigr\|_2^2
\Bigr)
\,\mathrm{d}\mu_k(x)
\right),
\qquad
\beta_c>0.
\label{eq:pics-tail-energy}
\end{equation}
Eq.~\eqref{eq:pics-mean-energy} controls the average residual content, whereas Eq.~\eqref{eq:pics-tail-energy} supplies a smooth proxy for worst-point risk through a log-sum-exp geometry. For the boundary and interface channels, we write
\begin{equation}
\mathcal{B}_c(\theta)
=
\int_{\partial\Omega_c}
\bigl\|
\widehat{\mathbf b}_{c,\theta}(x)
\bigr\|_2^2
\,\mathrm{d}\sigma_c(x),
\label{eq:pics-boundary-consistency}
\end{equation}
\begin{equation}
\mathcal{I}_c(\theta;\mu_k)
=
\int_{\Gamma_c}
\bigl\|
\widehat{\mathbf i}_{c,\theta}(x)
\bigr\|_2^2
\,\mathrm{d}\mu_k^{\Gamma}(x),
\label{eq:pics-interface-consistency}
\end{equation}
where \(\widehat{\mathbf b}_{c,\theta}\) and \(\widehat{\mathbf i}_{c,\theta}\) collect the normalized boundary and interface mismatch channels, and \(\mu_k^{\Gamma}\) denotes the interface trace measure induced by the current empirical sampling rule. When \(\Gamma_c=\varnothing\), we set \(\mathcal{I}_c(\theta;\mu_k)=0\). Detailed bounds and proofs for Eqs.~\eqref{eq:pics-composite-energy}--\eqref{eq:pics-interface-consistency} are deferred to Sections~V and VIII of the SM. From the same normalized residual section, PICS resolves a certificate field
\begin{equation}
\mathfrak{C}_{c,\theta}(x)
=
\bigl\|
\widehat{\mathbf r}_{c,\theta}(x)
\bigr\|_{\infty}
=
\max_{1\le m\le M_c}
\left|
\widehat r_{c,\theta,m}(x)
\right|,
\label{eq:pics-certificate-field}
\end{equation}
so PICS does not rely solely on mean residual descent; it explicitly resolves a certificate field that identifies fieldwise and global extreme regions. Section~VI of the SM gives the detailed extraction of extreme sets, persistent memory, and case-wise certified regions. The training stage then acts on a transported empirical measure,
\begin{equation}
\mu_k^{\mathrm{train}}
=
\rho\,\mu_k^{\mathrm{mem}}
+
(1-\rho)\,\mu_k^{\mathrm{reg}},
\qquad
0\le \rho\le 1,
\label{eq:pics-training-measure}
\end{equation}
where \(\mu_k^{\mathrm{mem}}\) denotes the risk-memory component and \(\mu_k^{\mathrm{reg}}\) denotes the regular exploration pool. We further write
\begin{equation}
\mu_k^{\mathrm{reg}}
=
(1-\eta_k)\,\mu_c^{\mathrm{base}}
+
\eta_k\,\mu_{k,c}^{\mathrm{ext}},
\qquad
0\le \eta_k\le 1,
\label{eq:pics-regular-pool}
\end{equation}
where \(\mu_c^{\mathrm{base}}\) denotes the case-dependent base measure and \(\mu_{k,c}^{\mathrm{ext}}\) enriches the regular pool by certified extreme regions and case-dependent geometry-active regions. Sections~VI, VII, and IX of the SM give the transport dynamics, persistent memory update, and parameter rules.

These objects close the solver loop in Fig.~\ref{F1}. At iteration \(k\), PICS moves from the admissible state to the induced fields, from the induced fields to the restricted jet, from the restricted jet to the normalized residual section, from the residual section to the composite residual geometry and certificate field, and from the certificate field to the transported training measure before advancing the parameter state according to
\begin{equation}
\theta_{k+1}
=
\mathcal{U}_k\!\left(\theta_k;\mu_k^{\mathrm{train}}\right),
\label{eq:pics-solver-update}
\end{equation}
where \(\mathcal{U}_k\) denotes the update map induced by certificate-governed descent dynamics on the transported empirical measure. PICS therefore defines a closed solver framework on admissible states, restricted jets, residual geometry, certificate fields, and empirical measure transport. The detailed definitions, derivations, propositions, proofs, and parameter analyses are given in the SM.
\subsection*{Aggregate quantitative comparison}

Before turning to the case-resolved visual evidence in Figs.~\ref{F2}--\ref{F4}, we first compare PICS, PINN, DGM, and DRM at the aggregate benchmark level through Tables~\ref{T1}--\ref{T2}. This comparison does not restate the solver construction summarized in Fig.~\ref{F1}; instead, it tests whether the structured solver objects translate into a quantitatively reliable benchmark profile across the three cases and the five coupled fields \((u,v,p,\phi,T)\). Table~\ref{T1} therefore serves as an aggregate fieldwise benchmark matrix, while Table~\ref{T2} reports the corresponding computational cost. Read together, the two tables answer the benchmark-level question that precedes any case-by-case discussion: whether PICS improves multi-field accuracy in a consistent way and whether that gain remains computationally practical.

\begin{table*}[t]
\centering
\footnotesize
\setlength{\tabcolsep}{4pt}
\renewcommand{\arraystretch}{1.12}
\caption{Aggregate fieldwise error metrics of PICS, PINN, DGM, and DRM across the three benchmark cases. For each case and each physical field, the reported quantities are the root-mean-square error (RMSE), mean-square error (MSE), mean absolute error (MAE), and relative \(L_2\) error (relL2). Panel (a) reports RMSE and MAE, while panel (b) reports MSE and relL2.}
\label{T1}

\textbf{(a) RMSE and MAE}

\vspace{2pt}

\begin{tabular}{llcccccccc}
\toprule
\multicolumn{2}{c}{\textbf{Benchmark}} &
\multicolumn{2}{c}{\textbf{PICS}} &
\multicolumn{2}{c}{\textbf{PINN}} &
\multicolumn{2}{c}{\textbf{DGM}} &
\multicolumn{2}{c}{\textbf{DRM}} \\
\cmidrule(lr){1-2}
\cmidrule(lr){3-4}
\cmidrule(lr){5-6}
\cmidrule(lr){7-8}
\cmidrule(lr){9-10}
\textbf{Case} & \textbf{Field}
& \textbf{RMSE} & \textbf{MAE}
& \textbf{RMSE} & \textbf{MAE}
& \textbf{RMSE} & \textbf{MAE}
& \textbf{RMSE} & \textbf{MAE} \\
\midrule

\textbf{Case 1} & $u$
& $\bm{9.5304\mathrm{e}{-02}}$ & $\bm{6.1297\mathrm{e}{-02}}$
& 4.4855e-01 & 3.4835e-01
& 3.7898e-01 & 2.9425e-01
& 2.0924e-01 & 1.6563e-01 \\
& $v$
& $\bm{1.0457\mathrm{e}{-01}}$ & $\bm{6.9456\mathrm{e}{-02}}$
& 6.9973e-01 & 5.3079e-01
& 1.3181e+00 & 6.6802e-01
& 3.7721e-01 & 2.0071e-01 \\
& $p$
& $\bm{5.6530\mathrm{e}{-02}}$ & $\bm{4.1064\mathrm{e}{-02}}$
& 5.3141e-01 & 4.1172e-01
& 6.3554e-01 & 3.8220e-01
& 2.7062e-01 & 1.8126e-01 \\
& $\phi$
& $\bm{1.8304\mathrm{e}{-02}}$ & $\bm{1.3958\mathrm{e}{-02}}$
& 4.7131e-02 & 3.3461e-02
& 6.1321e-02 & 5.3469e-02
& 4.2133e-02 & 3.2803e-02 \\
& $T$
& $\bm{1.2508\mathrm{e}{+00}}$ & $\bm{7.8159\mathrm{e}{-01}}$
& 3.8626e+00 & 3.0190e+00
& 8.3923e+00 & 5.8643e+00
& 2.6163e+00 & 1.7698e+00 \\

\addlinespace[2pt]
\textbf{Case 2} & $u$
& $\bm{1.9282\mathrm{e}{-01}}$ & $\bm{1.3381\mathrm{e}{-01}}$
& 3.9869e-01 & 3.0912e-01
& 3.6125e-01 & 2.6655e-01
& 4.3488e-01 & 2.9218e-01 \\
& $v$
& $\bm{3.6768\mathrm{e}{-01}}$ & $\bm{2.3461\mathrm{e}{-01}}$
& 1.0066e+00 & 5.9017e-01
& 8.8202e-01 & 3.9773e-01
& 1.2211e+00 & 5.9815e-01 \\
& $p$
& $\bm{6.4742\mathrm{e}{-02}}$ & $\bm{4.4389\mathrm{e}{-02}}$
& 4.4610e-01 & 3.7119e-01
& 4.5830e-01 & 3.2431e-01
& 4.5918e-01 & 3.2850e-01 \\
& $\phi$
& 5.9323e-02 & 4.2843e-02
& 7.0862e-02 & 5.8512e-02
& $\bm{3.5257\mathrm{e}{-02}}$ & 2.9914e-02
& 3.7860e-02 & $\bm{2.7487\mathrm{e}{-02}}$ \\
& $T$
& $\bm{6.7047\mathrm{e}{+00}}$ & $\bm{5.0887\mathrm{e}{+00}}$
& 9.3198e+00 & 6.2174e+00
& 2.0895e+01 & 1.5830e+01
& 1.4178e+01 & 9.9819e+00 \\

\addlinespace[2pt]
\textbf{Case 3} & $u$
& 1.0413e-01 & 7.7971e-02
& 1.0798e-01 & 8.1738e-02
& $\bm{4.7499\mathrm{e}{-02}}$ & $\bm{3.6858\mathrm{e}{-02}}$
& 7.3736e-01 & 4.4248e-01 \\
& $v$
& 1.1960e-01 & 8.1906e-02
& 1.1902e-01 & 8.7381e-02
& $\bm{3.9974\mathrm{e}{-02}}$ & $\bm{2.9422\mathrm{e}{-02}}$
& 7.5210e-01 & 3.8060e-01 \\
& $p$
& 5.5176e-02 & 3.7077e-02
& 1.1346e-01 & 9.0672e-02
& $\bm{2.9630\mathrm{e}{-02}}$ & $\bm{2.2398\mathrm{e}{-02}}$
& 3.3005e-01 & 2.4075e-01 \\
& $\phi$
& 6.5500e-02 & 5.2603e-02
& 1.5796e-02 & 1.0140e-02
& $\bm{4.5293\mathrm{e}{-03}}$ & $\bm{2.2102\mathrm{e}{-03}}$
& 2.6246e-02 & 2.1032e-02 \\
& $T$
& 2.3978e+00 & 1.8088e+00
& 1.9655e+00 & 1.4166e+00
& $\bm{5.8653\mathrm{e}{-01}}$ & $\bm{4.3372\mathrm{e}{-01}}$
& 5.4928e+00 & 3.2555e+00 \\

\bottomrule
\end{tabular}

\vspace{6pt}

\textbf{(b) MSE and relL2}

\vspace{2pt}

\begin{tabular}{llcccccccc}
\toprule
\multicolumn{2}{c}{\textbf{Benchmark}} &
\multicolumn{2}{c}{\textbf{PICS}} &
\multicolumn{2}{c}{\textbf{PINN}} &
\multicolumn{2}{c}{\textbf{DGM}} &
\multicolumn{2}{c}{\textbf{DRM}} \\
\cmidrule(lr){1-2}
\cmidrule(lr){3-4}
\cmidrule(lr){5-6}
\cmidrule(lr){7-8}
\cmidrule(lr){9-10}
\textbf{Case} & \textbf{Field}
& \textbf{MSE} & \textbf{relL2}
& \textbf{MSE} & \textbf{relL2}
& \textbf{MSE} & \textbf{relL2}
& \textbf{MSE} & \textbf{relL2} \\
\midrule

\textbf{Case 1} & $u$
& $\bm{9.0828\mathrm{e}{-03}}$ & $\bm{1.8617\mathrm{e}{-01}}$
& 2.0120e-01 & 8.7623e-01
& 1.4362e-01 & 7.4032e-01
& 4.3781e-02 & 4.0874e-01 \\
& $v$
& $\bm{1.0934\mathrm{e}{-02}}$ & $\bm{9.0172\mathrm{e}{-02}}$
& 4.8962e-01 & 6.0340e-01
& 1.7374e+00 & 1.1366e+00
& 1.4229e-01 & 3.2528e-01 \\
& $p$
& $\bm{3.1956\mathrm{e}{-03}}$ & $\bm{5.4226\mathrm{e}{-02}}$
& 2.8240e-01 & 5.0975e-01
& 4.0392e-01 & 6.0964e-01
& 7.3237e-02 & 2.5959e-01 \\
& $\phi$
& $\bm{3.3503\mathrm{e}{-04}}$ & $\bm{1.0867\mathrm{e}{-01}}$
& 2.2214e-03 & 2.7982e-01
& 3.7602e-03 & 3.6406e-01
& 1.7752e-03 & 2.5014e-01 \\
& $T$
& $\bm{1.5644\mathrm{e}{+00}}$ & $\bm{4.1603\mathrm{e}{-03}}$
& 1.4920e+01 & 1.2848e-02
& 7.0430e+01 & 2.7915e-02
& 6.8448e+00 & 8.7023e-03 \\

\addlinespace[2pt]
\textbf{Case 2} & $u$
& $\bm{3.7179\mathrm{e}{-02}}$ & $\bm{3.8236\mathrm{e}{-01}}$
& 1.5896e-01 & 7.9062e-01
& 1.3050e-01 & 7.1636e-01
& 1.8912e-01 & 8.6238e-01 \\
& $v$
& $\bm{1.3519\mathrm{e}{-01}}$ & $\bm{3.5842\mathrm{e}{-01}}$
& 1.0132e+00 & 9.8121e-01
& 7.7797e-01 & 8.5981e-01
& 1.4911e+00 & 1.1904e+00 \\
& $p$
& $\bm{4.1915\mathrm{e}{-03}}$ & $\bm{6.3014\mathrm{e}{-02}}$
& 1.9900e-01 & 4.3419e-01
& 2.1004e-01 & 4.4606e-01
& 2.1084e-01 & 4.4692e-01 \\
& $\phi$
& 3.5192e-03 & 4.8744e-01
& 5.0214e-03 & 5.8225e-01
& $\bm{1.2431\mathrm{e}{-03}}$ & $\bm{2.8970\mathrm{e}{-01}}$
& 1.4334e-03 & 3.1108e-01 \\
& $T$
& $\bm{4.4953\mathrm{e}{+01}}$ & $\bm{2.2329\mathrm{e}{-02}}$
& 8.6858e+01 & 3.1038e-02
& 4.3659e+02 & 6.9586e-02
& 2.0101e+02 & 4.7216e-02 \\

\addlinespace[2pt]
\textbf{Case 3} & $u$
& 1.0843e-02 & 1.7591e-01
& 1.1660e-02 & 1.8242e-01
& $\bm{2.2561\mathrm{e}{-03}}$ & $\bm{8.0243\mathrm{e}{-02}}$
& 5.4370e-01 & 1.2457e+00 \\
& $v$
& 1.4304e-02 & 2.4282e-01
& 1.4165e-02 & 2.4163e-01
& $\bm{1.5979\mathrm{e}{-03}}$ & $\bm{8.1155\mathrm{e}{-02}}$
& 5.6565e-01 & 1.5269e+00 \\
& $p$
& 3.0444e-03 & 5.4713e-02
& 1.2873e-02 & 1.1251e-01
& $\bm{8.7795\mathrm{e}{-04}}$ & $\bm{2.9382\mathrm{e}{-02}}$
& 1.0893e-01 & 3.2728e-01 \\
& $\phi$
& 4.2903e-03 & 4.5846e-01
& 2.4950e-04 & 1.1056e-01
& $\bm{2.0514\mathrm{e}{-05}}$ & $\bm{3.1702\mathrm{e}{-02}}$
& 6.8884e-04 & 1.8370e-01 \\
& $T$
& 5.7496e+00 & 7.9945e-03
& 3.8631e+00 & 6.5530e-03
& $\bm{3.4401\mathrm{e}{-01}}$ & $\bm{1.9555\mathrm{e}{-03}}$
& 3.0171e+01 & 1.8313e-02 \\

\bottomrule
\end{tabular}
\end{table*}

The main message of Table~\ref{T1} is aggregate balance rather than pointwise dominance. In a coupled PDE benchmark, a method may remain locally competitive on a particular field or metric and still be less reliable overall if its error profile fluctuates sharply across fields, metrics, or cases. Table~\ref{T1} should therefore be read both row-wise and column-wise: row-wise, it reports the case--field behavior of each method under RMSE, MSE, MAE, and relL2; column-wise, it shows whether a method sustains its accuracy across heterogeneous fields rather than improving one component at the expense of another. Under this reading, the relevant advantage of PICS lies in aggregate fieldwise balance, reduced cross-case drift, and a more stable error profile across the four reported metrics. Figs.~\ref{F2}--\ref{F4} then resolve how these aggregate trends appear in space through the predicted fields, the maximum-error maps, and the case-wise solver diagnostics.

\begin{table}[t]
\centering
\scriptsize
\caption{Aggregate runtime comparison of PICS, PINN, DGM, and DRM across the three benchmark cases. The reported values are total wall-clock runtime in seconds. More detailed runtime decomposition is provided in Section IX and Section X of the SM.}
\label{T2}
\begin{tabular*}{\columnwidth}{@{\extracolsep{\fill}}lcccc@{}}
\toprule
\textbf{Case} & \textbf{PICS (s)} & \textbf{PINN (s)} & \textbf{DGM (s)} & \textbf{DRM (s)} \\
\midrule
\textbf{Case 1} & 17445.64 & 7385.76  & 28146.37 & 6661.18  \\
\textbf{Case 2} & 21154.34 & 10369.48 & 34563.25 & 10912.70 \\
\textbf{Case 3} & 22389.67 & 12700.58 & 36750.74 & 12420.56 \\
\bottomrule
\end{tabular*}
\end{table}

Table~\ref{T2} places the accuracy gains in computational context. Runtime is not the primary selling point of PICS, but the accuracy improvement is not obtained at an impractical cost. Across the three cases, the total runtime remains operationally reasonable at the present benchmark scale and substantially below DGM, even though PICS is not uniformly the fastest method. The defensible statement is therefore not speed dominance, but that the aggregate accuracy gain of PICS is achieved without a prohibitive runtime penalty. This benchmark-level reading of Tables~\ref{T1}--\ref{T2} sets the stage for the case-specific analyses in Figs.~\ref{F2}--\ref{F4}.
\subsection*{Case 1: canonical coupled interface-type recovery}
\begin{figure*}[p]
    \centering
    \makebox[\textwidth][c]{%
        \includegraphics[width=1.50\textwidth,trim=7mm 0mm 4mm 0mm,clip]{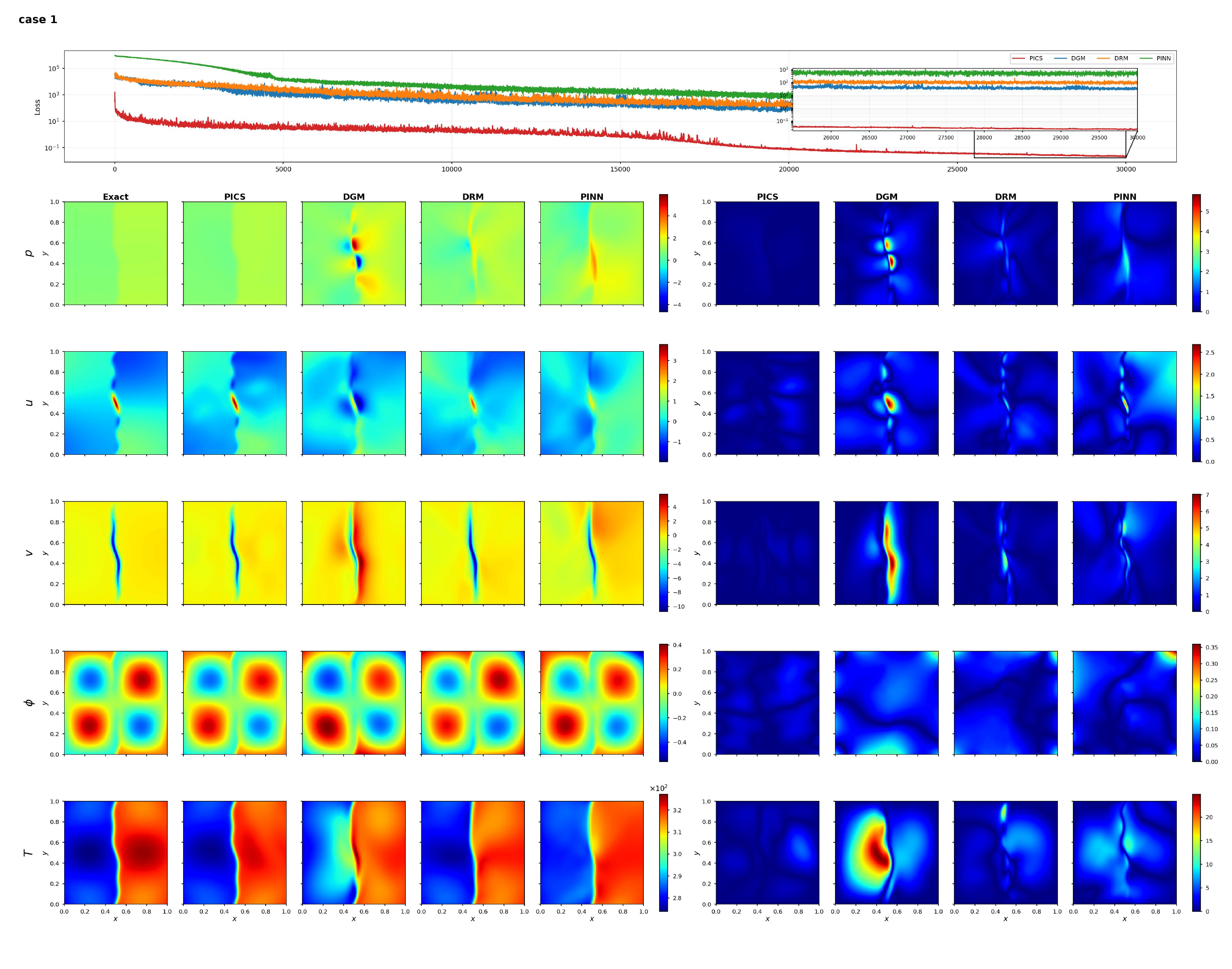}
    }
    \caption{\textbf{The first benchmark realization of the proposed PICS solver.} The top trajectory panel records the evolution of the residual-energy and certificate-governed solver state during training. The field-reconstruction panels compare the predicted \(p\), \(u\), \(v\), \(\phi\), and \(T\) fields produced by PICS, PINN, DGM, and DRM against the analytic reference under the first benchmark configuration. The corresponding maximum-error maps show that PICS yields more balanced cross-field recovery and contracts the dominant error hotspots toward the interface-active region, whereas the baseline solvers exhibit broader spreading, mild amplitude drift, or stronger interface-localized distortion.}
    \label{F2}
\end{figure*}
Case 1 is the first canonical coupled benchmark in this work. Its role is to provide a geometry-active, cross-field coupled, analytically tractable testbed for solver verification. It does not aim to reproduce a specific experimental platform. Instead, it isolates whether interface-active geometry, cross-field coupling, representation-level admissibility, and certificate-aware transport can be recovered in a controlled benchmark setting with known truth. For this reason, the governing PDE system is first specified, after which an analytic reference state admitted by that system is introduced.

On the unit square \(\Omega_1=(0,1)^2\), the benchmark PDE system is
\begin{equation}
u\,\partial_x u
+
v\,\partial_y u
+
\partial_x p
-
\nu\,(\partial_{xx}u+\partial_{yy}u)
-
\chi\,\rho_e\,\partial_x\phi
=
f_u
\qquad \text{in } \Omega_1,
\label{eq:case1-u-momentum}
\end{equation}
\begin{equation}
u\,\partial_x v
+
v\,\partial_y v
+
\partial_y p
-
\nu\,(\partial_{xx}v+\partial_{yy}v)
-
\chi\,\rho_e\,\partial_y\phi
=
f_v
\qquad \text{in } \Omega_1,
\label{eq:case1-v-momentum}
\end{equation}
\begin{equation}
-\varepsilon\,(\partial_{xx}\phi+\partial_{yy}\phi)
=
\rho_e
\qquad \text{in } \Omega_1,
\label{eq:case1-electrostatic}
\end{equation}
\begin{equation}
u\,\partial_x T
+
v\,\partial_y T
-
\kappa\,(\partial_{xx}T+\partial_{yy}T)
-
\sigma_J\!\left[(\partial_x\phi)^2+(\partial_y\phi)^2\right]
=
f_T
\qquad \text{in } \Omega_1,
\label{eq:case1-thermal}
\end{equation}
where \(\nu\), \(\varepsilon\), \(\chi\), \(\kappa\), and \(\sigma_J\) denote the viscosity, permittivity, electrostatic coupling strength, thermal diffusivity, and Joule-heating coefficient, respectively. Incompressibility is enforced structurally through Eq.~\eqref{eq:pics-incompressibility} and is therefore not re-imposed as an additional residual channel.

Case 1 is organized around a single interface-active transition geometry, which makes it the canonical benchmark for testing whether the solver can recover coupled fields in a geometry-sensitive yet structurally admissible manner. We define the interface profile by
\begin{equation}
x_{\Gamma}(y)
=
0.5+0.03\sin(2\pi y)+0.01\sin(6\pi y),
\label{eq:case1-interface-profile}
\end{equation}
and the corresponding transition-layer function by
\begin{equation}
s(x,y)
=
\tanh\!\left(\frac{x-x_{\Gamma}(y)}{\Delta}\right),
\qquad
\Delta=0.025.
\label{eq:case1-transition-layer}
\end{equation}
The geometric role of this interface profile within the gate-structured manifold is detailed in Section II of the SM.

The benchmark PDE system above admits the following analytic reference state:
\begin{equation}
\psi^{\ast}(x,y)
=
0.12\sin(\pi x)\sin(\pi y)
+
0.035\,s(x,y)\sin(2\pi y)
+
0.015\cos(2\pi x)\sin(\pi y),
\label{eq:case1-psi-star}
\end{equation}
\begin{equation}
p^{\ast}(x,y)
=
0.18\cos(\pi x)\cos(\pi y)
+
0.06\,s(x,y)
+
0.025\sin(2\pi x)\sin(2\pi y),
\label{eq:case1-p-star}
\end{equation}
\begin{equation}
\phi^{\ast}(x,y)
=
0.20\sin(2\pi x)\sin(\pi y)
+
0.07\,s(x,y)\cos(\pi y)
+
0.03\sin(3\pi x)\sin(2\pi y),
\label{eq:case1-phi-star}
\end{equation}
\begin{equation}
T^{\ast}(x,y)
=
1.00
+
0.22\,s(x,y)
+
0.08\cos(\pi x)\sin(\pi y)
+
0.05\sin(2\pi x)\sin(2\pi y).
\label{eq:case1-T-star}
\end{equation}
The reference state combines a localized interface-sensitive layer with smooth global harmonics, so that the benchmark probes both geometry-active recovery and global field coherence. By the induced-field construction already introduced in Eq.~\eqref{eq:pics-field-induction}, the benchmark velocity pair is obtained from the reference streamfunction as
\begin{equation}
u^{\ast}(x,y)=\partial_y\psi^{\ast}(x,y),
\qquad
v^{\ast}(x,y)=-\partial_x\psi^{\ast}(x,y).
\label{eq:case1-velocity-star}
\end{equation}
Section III of the SM records the explicit derivative expansions, and Section VIII of the SM records the corresponding structural consistency statements.

Substituting the analytic reference state from Eqs.~\eqref{eq:case1-psi-star}--\eqref{eq:case1-T-star} into the component-wise Case 1 benchmark system yields the corresponding charge density and source-consistent forcing terms:
\begin{equation}
\rho_e^{\ast}
=
-\varepsilon\,(\partial_{xx}\phi^{\ast}+\partial_{yy}\phi^{\ast}),
\label{eq:case1-rho-star}
\end{equation}
\begin{equation}
f_u^{\ast}
=
u^{\ast}\partial_x u^{\ast}
+
v^{\ast}\partial_y u^{\ast}
+
\partial_x p^{\ast}
-
\nu\,(\partial_{xx}u^{\ast}+\partial_{yy}u^{\ast})
-
\chi\,\rho_e^{\ast}\partial_x\phi^{\ast},
\label{eq:case1-fu-star}
\end{equation}
\begin{equation}
f_v^{\ast}
=
u^{\ast}\partial_x v^{\ast}
+
v^{\ast}\partial_y v^{\ast}
+
\partial_y p^{\ast}
-
\nu\,(\partial_{xx}v^{\ast}+\partial_{yy}v^{\ast})
-
\chi\,\rho_e^{\ast}\partial_y\phi^{\ast},
\label{eq:case1-fv-star}
\end{equation}
\begin{equation}
f_T^{\ast}
=
u^{\ast}\partial_x T^{\ast}
+
v^{\ast}\partial_y T^{\ast}
-
\kappa\,(\partial_{xx}T^{\ast}+\partial_{yy}T^{\ast})
-
\sigma_J\!\left[(\partial_x\phi^{\ast})^2+(\partial_y\phi^{\ast})^2\right].
\label{eq:case1-fT-star}
\end{equation}
These forcing terms are not externally prescribed from an experimental platform; instead, they are inferred from the analytic reference state so that the benchmark remains source-consistent and analytically verifiable. The full differential expansions are given in Section IV and Section V of the SM.

Fig.~\ref{F2} provides the first spatially resolved realization of the solver loop summarized in Fig.~\ref{F1}. In the field-reconstruction panels, the comparison centers on whether a method can recover \(p\), \(u\), \(v\), \(\phi\), and \(T\) with coherent geometry-sensitive structure rather than on whether it fits an isolated smooth component slightly better than its competitors. Under this canonical interface-sensitive coupled benchmark, PICS preserves the interface-active structure and the smooth global background more coherently across the full field set, whereas PINN, DGM, and DRM exhibit more pronounced interface-localized distortion, mild amplitude drift, or broader smearing in the transition band. This is the correct case-level interpretation of Table~\ref{T1}: the advantage of PICS in Case 1 lies in more stable multi-field behavior, not in entrywise dominance.

The max-error maps in Fig.~\ref{F2} sharpen that comparison in space. Under PICS, the dominant hotspots contract toward the interface-active zone and remain weaker and more localized than those of the baselines. This pattern is consistent with the solver objects summarized in Fig.~\ref{F1}: the normalized residual section, the composite residual geometry, and the certificate field not only regulate mean discrepancy, but also suppress persistent high-risk regions more effectively. The spatial error structure in Fig.~\ref{F2} therefore resolves, at field level, the aggregate trend reported in Table~\ref{T1}.

The trajectory panel at the top of Fig.~\ref{F2} completes the interpretation. It records the evolution of the residual-energy and certificate-guided solver state associated with the solver loop in Fig.~\ref{F1}. Together with Table~\ref{T2}, Fig.~\ref{F2} shows that the Case 1 gain is achieved at practical computational cost. Section IX of the SM gives the complete parameter analysis for this case.

\subsection*{Case 2: thermo-viscous Leray-regularized coupled transport}
\begin{figure*}[p]
    \centering
    \makebox[\textwidth][c]{%
        \includegraphics[width=1.50\textwidth,trim=7mm 0mm 4mm 0mm,clip]{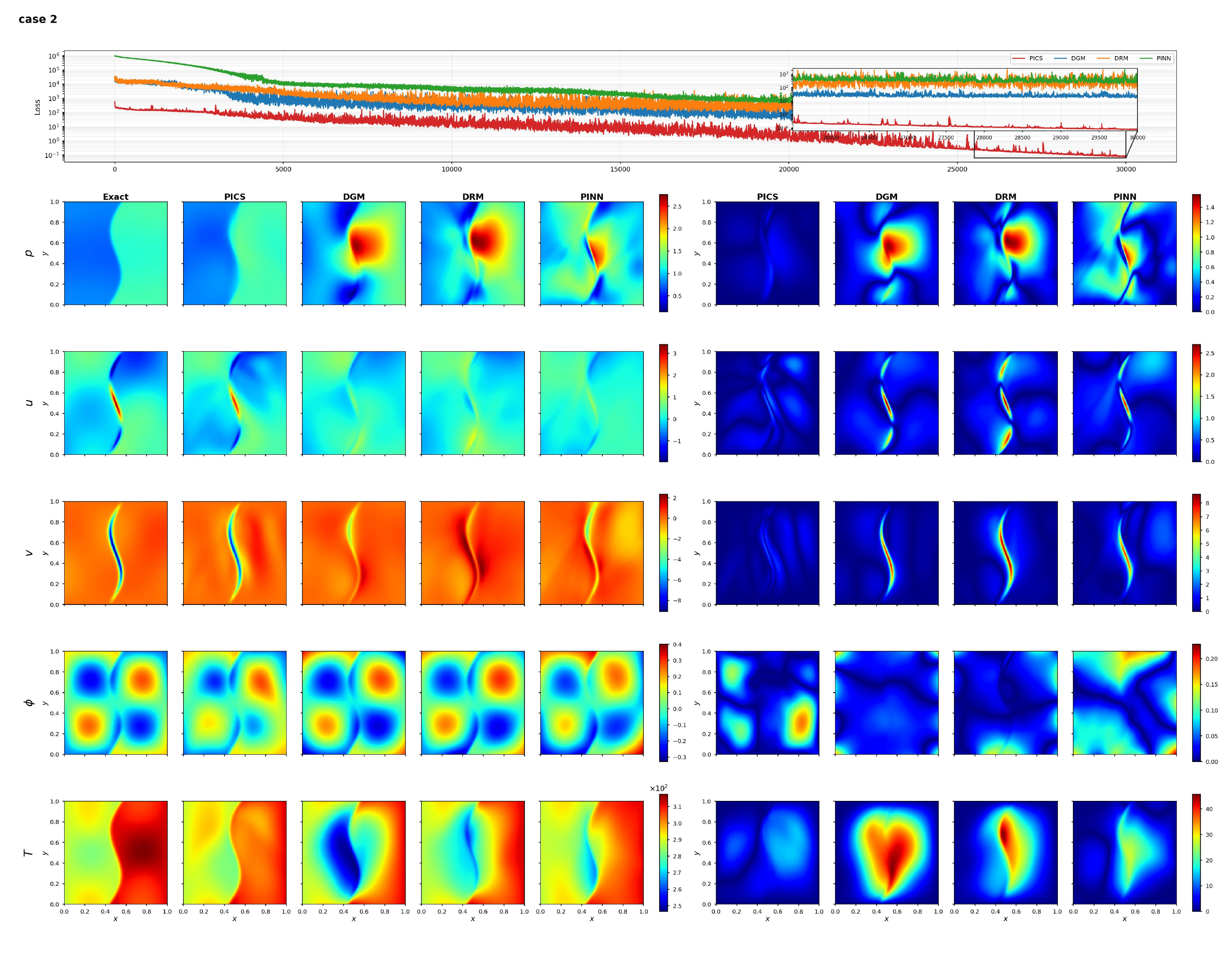}
    }
   \caption{\textbf{The second benchmark realization of the proposed PICS solver under altered closure structure.} The top trajectory panel shows the evolution of the residual energy and the certificate-guided solver state under the Case 2 thermo-viscous and Leray-regularized closure. The field panels compare the predicted \(p\), \(u\), \(v\), \(\phi\), and \(T\) fields from PICS, PINN, DGM, and DRM with the analytic reference for the second benchmark configuration. The maximum-error maps show that PICS preserves more stable cross-field coherence and keeps the dominant error concentrations weaker and more localized.}
    \label{F3}
\end{figure*}
Case 2 is the first closure-modified benchmark in this work. Its role is to test whether the advantage of PICS survives a genuine change in the transport-diffusion law, rather than merely a change of coefficients within the same canonical setting. While constructed as an analytically tractable benchmark, this regime captures the core mathematical difficulties of subgrid-scale turbulence modeling (via Leray regularization) and non-isothermal complex flows (via thermo-viscous diffusion). Instead of merely testing an arbitrary mathematical alteration, it probes solver behavior under physically motivated closure modifications where geometry-active transport and strong cross-field coupling act simultaneously. For this reason, the governing PDE system is first specified, after which an analytic reference state admitted by that system is introduced.

On the unit square \(\Omega_2=(0,1)^2\), we consider the coupled benchmark system
\begin{equation}
\bar u\,\partial_x u
+
\bar v\,\partial_y u
+
\partial_x p
-
\partial_x\!\bigl(\nu_T\,\partial_x u\bigr)
-
\partial_y\!\bigl(\nu_T\,\partial_y u\bigr)
+
\alpha_E\,\rho_e\,\partial_x\phi
=
f_u
\qquad \text{in } \Omega_2,
\label{eq:case2-u-momentum}
\end{equation}
\begin{equation}
\bar u\,\partial_x v
+
\bar v\,\partial_y v
+
\partial_y p
-
\partial_x\!\bigl(\nu_T\,\partial_x v\bigr)
-
\partial_y\!\bigl(\nu_T\,\partial_y v\bigr)
+
\alpha_E\,\rho_e\,\partial_y\phi
=
f_v
\qquad \text{in } \Omega_2,
\label{eq:case2-v-momentum}
\end{equation}
\begin{equation}
-\varepsilon\,(\partial_{xx}\phi+\partial_{yy}\phi)
=
\rho_e
\qquad \text{in } \Omega_2,
\label{eq:case2-electrostatic}
\end{equation}
\begin{equation}
\bar u\,\partial_x T
+
\bar v\,\partial_y T
-
\kappa\,(\partial_{xx}T+\partial_{yy}T)
-
\sigma_J\!\left[(\partial_x\phi)^2+(\partial_y\phi)^2\right]
=
f_T
\qquad \text{in } \Omega_2,
\label{eq:case2-thermal}
\end{equation}
where \(\bar u\) and \(\bar v\) denote the regularized transport pair, \(\nu_T\) denotes the temperature-dependent diffusion coefficient, \(\alpha_E\) denotes the electrostatic coupling strength, \(\varepsilon\) denotes the permittivity, \(\kappa\) denotes the thermal diffusivity, and \(\sigma_J\) denotes the Joule-heating coefficient. Incompressibility is enforced structurally through Eq.~\eqref{eq:pics-incompressibility} and is therefore not re-imposed as an additional residual channel.

Case 2 involves a transport-active layer geometry that enters both the benchmark fields and the geometry-aware enrichment mechanism. We define the curved layer profile by
\begin{equation}
x_s(y)
=
0.50+0.12\,(y-0.50)+0.08\sin(2\pi y),
\label{eq:case2-layer-profile}
\end{equation}
and the corresponding transition variable by
\begin{equation}
s(x,y)
=
\tanh\!\left(\frac{x-x_s(y)}{\Delta_s}\right),
\qquad
\Delta_s=0.018.
\label{eq:case2-layer-indicator}
\end{equation}
This layer geometry makes Case 2 fundamentally different from Case 1: it does not merely perturb a canonical interface, but instead injects a curved transport-active structure into the closure itself. Section II of the SM gives the geometric role of this layer within the gate-structured manifold.

The benchmark PDE system above admits the following analytic reference state:
\begin{equation}
\begin{aligned}
    \psi^{\ast}(x,y)
&=
0.10\sin(\pi x)\sin(\pi y)
+
0.16\,s(x,y)\sin(\pi y)
\\&+
0.05\sin(2\pi x)\sin(2\pi y)
+
0.02\cos(3\pi x)\sin(\pi y),
\end{aligned}
\label{eq:case2-psi-star}
\end{equation}
\begin{equation}
p^{\ast}(x,y)
=
1.00
+
0.24\,s(x,y)
+
0.05\cos(2\pi y)
+
0.012\sin(2\pi x+0.6\pi y),
\label{eq:case2-p-star}
\end{equation}
\begin{equation}
\begin{aligned}
    \phi^{\ast}(x,y)
&=
0.36\sin(2\pi x)\sin(2\pi y)
+
0.16\,s(x,y)\cos(\pi y)
\\&+
0.02\sin(3\pi x)\sin(\pi y)
+
0.012\cos(2\pi x+0.6\pi y),
\end{aligned}
\label{eq:case2-phi-star}
\end{equation}
\begin{equation}
\begin{aligned}
    T^{\ast}(x,y)
&=
300.0
+
12.0\,s(x,y)
+
6.0\sin(2\pi x)\cos(2\pi y)
\\&+
3.0\exp\!\left[-\frac{(x-0.25)^2+(y-0.75)^2}{0.020}\right]
\\&-
2.5\exp\!\left[-\frac{(x-0.78)^2+(y-0.28)^2}{0.018}\right]
\\&+
0.6\sin(2\pi x+0.6\pi y),
\end{aligned}
\label{eq:case2-T-star}
\end{equation}
where the same geometry-sensitive layer enters several fields while the temperature field also drives the diffusion law itself. This construction makes Case 2 fundamentally different from Case 1: the layer affects not only the benchmark fields but also the closure itself through Leray-regularized transport and thermo-viscous diffusion. By the induced-field construction already introduced in Eq.~\eqref{eq:pics-field-induction}, the benchmark velocity pair is obtained from the reference streamfunction as
\begin{equation}
u^{\ast}(x,y)=\partial_y\psi^{\ast}(x,y),
\qquad
v^{\ast}(x,y)=-\partial_x\psi^{\ast}(x,y).
\label{eq:case2-velocity-star}
\end{equation}
Section III of the SM records the explicit derivative expansions, and Section VIII of the SM records the corresponding structural consistency statements.

Case 2 modifies the closure not only through the benchmark state but also through the transport and diffusion law itself. We define the Leray-regularized transport pair by
\begin{equation}
\bar u^{\ast}
=
u^{\ast}
+
\frac{L_s^2}{2}\,(\partial_{xx}u^{\ast}+\partial_{yy}u^{\ast}),
\qquad
\bar v^{\ast}
=
v^{\ast}
+
\frac{L_s^2}{2}\,(\partial_{xx}v^{\ast}+\partial_{yy}v^{\ast}),
\qquad
L_s=0.020,
\label{eq:case2-leray-velocity}
\end{equation}
and the thermo-viscous diffusion coefficient by
\begin{equation}
\begin{aligned}
    &\widetilde T^{\ast}
=
\frac{T^{\ast}-T_0}{T_1},
\quad
&\nu_T^{\ast}
=
\nu_0\exp(-A_T\widetilde T^{\ast}),
\quad
&T_0=300.0,\\ &T_1=20.0,\quad &\nu_0=0.010,\quad &A_T=0.80.
\end{aligned}
\label{eq:case2-thermoviscous}
\end{equation}
Sections IV and V of the SM give the detailed closure expansions associated with Eqs.~\eqref{eq:case2-leray-velocity}--\eqref{eq:case2-thermoviscous}.

Substituting the analytic reference state together with the regularized transport pair and the thermo-viscous coefficient into the component-wise Case 2 benchmark system yields the corresponding charge density and source-consistent forcing terms:
\begin{equation}
\rho_e^{\ast}
=
-\varepsilon\,(\partial_{xx}\phi^{\ast}+\partial_{yy}\phi^{\ast}),
\label{eq:case2-rho-star}
\end{equation}
\begin{equation}
f_u^{\ast}
=
\bar u^{\ast}\partial_x u^{\ast}
+
\bar v^{\ast}\partial_y u^{\ast}
+
\partial_x p^{\ast}
-
\partial_x\!\bigl(\nu_T^{\ast}\partial_x u^{\ast}\bigr)
-
\partial_y\!\bigl(\nu_T^{\ast}\partial_y u^{\ast}\bigr)
+
\alpha_E\,\rho_e^{\ast}\partial_x\phi^{\ast},
\label{eq:case2-fu-star}
\end{equation}
\begin{equation}
f_v^{\ast}
=
\bar u^{\ast}\partial_x v^{\ast}
+
\bar v^{\ast}\partial_y v^{\ast}
+
\partial_y p^{\ast}
-
\partial_x\!\bigl(\nu_T^{\ast}\partial_x v^{\ast}\bigr)
-
\partial_y\!\bigl(\nu_T^{\ast}\partial_y v^{\ast}\bigr)
+
\alpha_E\,\rho_e^{\ast}\partial_y\phi^{\ast},
\label{eq:case2-fv-star}
\end{equation}
\begin{equation}
f_T^{\ast}
=
\bar u^{\ast}\partial_x T^{\ast}
+
\bar v^{\ast}\partial_y T^{\ast}
-
\kappa\,(\partial_{xx}T^{\ast}+\partial_{yy}T^{\ast})
-
\sigma_J\!\left[(\partial_x\phi^{\ast})^2+(\partial_y\phi^{\ast})^2\right].
\label{eq:case2-fT-star}
\end{equation}
These forcing terms are not externally prescribed from an experimental platform; instead, they are inferred from the analytic reference state so that the benchmark remains source-consistent and analytically verifiable. The full differential expansions are given in Sections IV and V of the SM.

Fig.~\ref{F3} resolves the first benchmark in which the closure itself changes. In the field-reconstruction panels, the main issue is no longer whether a solver can track a single geometry-active interface under the canonical transport law, but whether it can maintain cross-field coherence once the layer geometry and the transport--diffusion law change at the same time. Under this altered regime, PICS keeps the curved layer, the smooth harmonic background, the electrostatic structure, and the temperature-linked modulation more coherent across \(u\), \(v\), \(p\), \(\phi\), and \(T\), whereas PINN, DGM, and DRM show broader layer-adjacent smearing, stronger local drift, or weaker multi-field consistency. This is the correct case-level interpretation of Table~\ref{T1}: the aggregate advantage of PICS persists when the closure itself is modified.

The max-error maps in Fig.~\ref{F3} sharpen that comparison in spatial terms. In Case 2, the most difficult regions cluster near the curved transport layer and around the thermal activity centers that feed back into \(\nu_T^{\ast}\). Under PICS, these error concentrations remain weaker and more localized, whereas under the baselines they spread more broadly along the transport-active layer and around the thermo-viscous activity zones. This spatial pattern matches the solver logic summarized in Fig.~\ref{F1}: the normalized residual section, the composite residual geometry, and the certificate field do not merely reduce aggregate discrepancy, but also suppress persistent high-risk regions more selectively.

The trajectory panel at the top of Fig.~\ref{F3} completes the interpretation. It records the evolution of the closed solver state under the modified Case 2 closure. Together with Table~\ref{T2}, Fig.~\ref{F3} shows that the Case 2 gain persists under altered closure while the runtime remains practical. Section IX of the SM gives the complete parameter analysis for this case.

\subsection*{Case 3: pressure-regularized screened electro-thermal coupling}
\begin{figure*}[p]
    \centering
    \makebox[\textwidth][c]{%
        \includegraphics[width=1.50\textwidth,trim=7mm 0mm 4mm 0mm,clip]{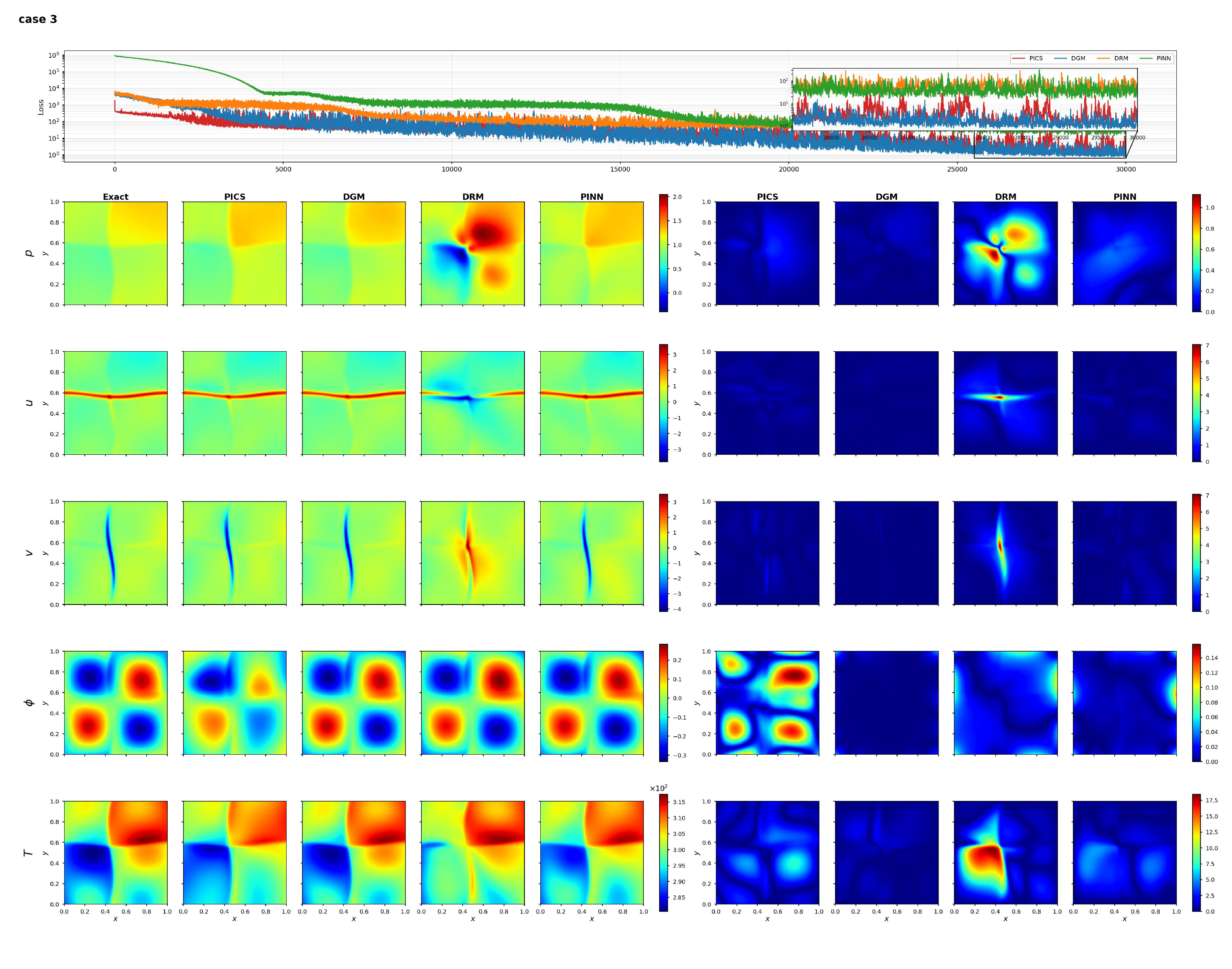}
    }
 \caption{\textbf{The third benchmark realization of the proposed PICS solver under the hardest closure configuration.} The top trajectory panel shows the evolution of the residual energy and the certificate-guided solver state for the most demanding benchmark in this work. The field panels compare the predicted \(p\), \(u\), \(v\), \(\phi\), and \(T\) fields from PICS, PINN, DGM, and DRM with the analytic reference for the pressure-regularized and screened electro-thermal closure of Case 3. The maximum-error maps show that PICS maintains more balanced cross-field recovery and confines the dominant error hotspots more effectively.}
    \label{F4}
\end{figure*}
Case 3 is the most demanding benchmark in this work. Its role is to test whether the aggregate advantage of PICS survives when the closure is further complicated by simultaneous pressure-gradient regularization and screened electrostatic coupling. Although formulated as a synthetic setting, this configuration mirrors the extreme structural demands of electrokinetic flows and singular perturbation limits in computational mechanics. Specifically, it probes solver robustness in a highly stringent regime where strict boundary-layer-like transitions (screened electrostatics via the Debye length), higher-order closure information (pressure regularization), geometry-sensitive fields, and cross-field coupling act simultaneously. For this reason, the governing PDE system is first specified, after which an analytic reference state admitted by that system is introduced.

On the unit square \(\Omega_3=(0,1)^2\), we consider the benchmark PDE system
\begin{equation}
u\,\partial_x u
+
v\,\partial_y u
+
\partial_x p
-
\ell_p^2\,\partial_x(\partial_{xx}p+\partial_{yy}p)
-
\nu\,(\partial_{xx}u+\partial_{yy}u)
-
\chi\,\rho_e\,\partial_x\phi
=
f_u
\qquad \text{in } \Omega_3,
\label{eq:case3-u-momentum}
\end{equation}
\begin{equation}
u\,\partial_x v
+
v\,\partial_y v
+
\partial_y p
-
\ell_p^2\,\partial_y(\partial_{xx}p+\partial_{yy}p)
-
\nu\,(\partial_{xx}v+\partial_{yy}v)
-
\chi\,\rho_e\,\partial_y\phi
=
f_v
\qquad \text{in } \Omega_3,
\label{eq:case3-v-momentum}
\end{equation}
\begin{equation}
-\varepsilon\,(\partial_{xx}\phi+\partial_{yy}\phi)
+
\kappa_D^2\,\phi
=
\rho_e
\qquad \text{in } \Omega_3,
\label{eq:case3-screened-electrostatic}
\end{equation}
\begin{equation}
u\,\partial_x T
+
v\,\partial_y T
-
\kappa\,(\partial_{xx}T+\partial_{yy}T)
-
\sigma_J\!\left[(\partial_x\phi)^2+(\partial_y\phi)^2\right]
=
f_T
\qquad \text{in } \Omega_3,
\label{eq:case3-thermal}
\end{equation}
where \(\ell_p\) denotes the pressure-gradient regularization length, \(\nu\) denotes the viscosity, \(\varepsilon\) denotes the permittivity, \(\kappa_D\) denotes the screening parameter, \(\chi\) denotes the electrostatic coupling strength, \(\kappa\) denotes the thermal diffusivity, and \(\sigma_J\) denotes the Joule-heating coefficient. Incompressibility is enforced structurally through Eq.~\eqref{eq:pics-incompressibility} and is therefore not re-imposed as an additional residual channel.

Case 3 involves a more complex geometry-sensitive modulation, consistent with the harder closure structure of the benchmark. We define the curved modulation profile by
\begin{equation}
x_r(y)
=
0.50
+
0.06\sin(2\pi y)
+
0.025\sin(6\pi y)
-
0.04\,(y-0.50)^2,
\label{eq:case3-geometry-profile}
\end{equation}
and the associated transition variable by
\begin{equation}
s(x,y)
=
\tanh\!\left(\frac{x-x_r(y)}{\Delta_r}\right),
\qquad
\Delta_r=0.016.
\label{eq:case3-geometry-indicator}
\end{equation}
This modulation is more demanding than the geometry used in Figs.~\ref{F2}--~\ref{F3}, because it enters a closure that already contains both a screened potential law and a pressure-regularized momentum balance. Section II of the SM gives the geometric role of this modulation within the gate-structured manifold.

The benchmark PDE system above admits the following analytic reference state:
\begin{equation}
\begin{aligned}
    \psi^{\ast}(x,y)
&=
0.11\sin(\pi x)\sin(\pi y)
+
0.055\sin(2\pi x)\sin(2\pi y)
\\&+
0.040\,s(x,y)\sin(\pi y)
+
0.018\cos(3\pi x)\sin(2\pi y),
\end{aligned}
\label{eq:case3-psi-star}
\end{equation}
\begin{equation}
\begin{aligned}
    p^{\ast}(x,y)
&=
0.22\cos(\pi x)\cos(\pi y)
+
0.075\,s(x,y)
+
0.030\sin(2\pi x)\sin(2\pi y)
\\&+
0.015\cos(3\pi x)\cos(\pi y)
+
0.010\exp\!\left[-\frac{(x-0.72)^2+(y-0.34)^2}{0.018}\right],
\end{aligned}
\label{eq:case3-p-star}
\end{equation}
\begin{equation}
\begin{aligned}
    \phi^{\ast}(x,y)
&=
0.24\sin(2\pi x)\sin(\pi y)
+
0.090\,s(x,y)\cos(\pi y)
\\&+
0.040\sin(3\pi x)\sin(2\pi y)
+
0.020\cos(2\pi x+0.8\pi y),
\end{aligned}
\label{eq:case3-phi-star}
\end{equation}
\begin{equation}
\begin{aligned}
    T^{\ast}(x,y)
&=
1.20
+
0.26\,s(x,y)
+
0.090\cos(\pi x)\sin(\pi y)
\\&+
0.060\sin(2\pi x)\sin(2\pi y)
+
0.030\exp\!\left[-\frac{(x-0.28)^2+(y-0.76)^2}{0.020}\right]
\\&-
0.022\exp\!\left[-\frac{(x-0.80)^2+(y-0.24)^2}{0.016}\right].
\end{aligned}
\label{eq:case3-T-star}
\end{equation}
The reference state is chosen so that the benchmark simultaneously probes geometry-active recovery, screened electrostatic coupling, and pressure-regularized closure sensitivity. By the induced-field construction already introduced in Eq.~\eqref{eq:pics-field-induction}, the benchmark velocity pair is obtained from the reference streamfunction as
\begin{equation}
u^{\ast}(x,y)=\partial_y\psi^{\ast}(x,y),
\qquad
v^{\ast}(x,y)=-\partial_x\psi^{\ast}(x,y).
\label{eq:case3-velocity-star}
\end{equation}
Section III of the SM records the explicit derivative expansions, and Section VIII of the SM records the corresponding structural consistency statements.

Case 3 modifies the closure through two simultaneous mechanisms: a pressure-gradient regularization acting on the momentum balance, and a screened electrostatic coupling acting on the charge-potential relation. The pressure-regularized gradient is written as
\begin{equation}
\nabla p^{\ast}_{\mathrm{reg}}
=
\nabla p^{\ast}
-
\ell_p^2\,\nabla(\partial_{xx}p^{\ast}+\partial_{yy}p^{\ast}),
\qquad
\ell_p=0.030,
\label{eq:case3-pressure-regularization}
\end{equation}
and the screened electrostatic relation satisfied by the reference potential is
\begin{equation}
-\varepsilon\,(\partial_{xx}\phi^{\ast}+\partial_{yy}\phi^{\ast})
+
\kappa_D^2\,\phi^{\ast}
=
\rho_e^{\ast},
\qquad
\varepsilon=1.0,
\qquad
\kappa_D=1.4.
\label{eq:case3-screened-reference}
\end{equation}
Sections IV and V of the SM give the full closure derivations associated with Eqs.~\eqref{eq:case3-pressure-regularization}--\eqref{eq:case3-screened-reference}.

Substituting the analytic reference state together with the pressure-regularized and screened closure terms into the component-wise Case 3 benchmark system yields the corresponding charge density and source-consistent forcing terms:
\begin{equation}
\rho_e^{\ast}
=
-\varepsilon\,(\partial_{xx}\phi^{\ast}+\partial_{yy}\phi^{\ast})
+
\kappa_D^2\,\phi^{\ast},
\label{eq:case3-rho-star}
\end{equation}
\begin{equation}
f_u^{\ast}
=
u^{\ast}\partial_x u^{\ast}
+
v^{\ast}\partial_y u^{\ast}
+
\partial_x p^{\ast}
-
\ell_p^2\,\partial_x(\partial_{xx}p^{\ast}+\partial_{yy}p^{\ast})
-
\nu\,(\partial_{xx}u^{\ast}+\partial_{yy}u^{\ast})
-
\chi\,\rho_e^{\ast}\partial_x\phi^{\ast},
\label{eq:case3-fu-star}
\end{equation}
\begin{equation}
f_v^{\ast}
=
u^{\ast}\partial_x v^{\ast}
+
v^{\ast}\partial_y v^{\ast}
+
\partial_y p^{\ast}
-
\ell_p^2\,\partial_y(\partial_{xx}p^{\ast}+\partial_{yy}p^{\ast})
-
\nu\,(\partial_{xx}v^{\ast}+\partial_{yy}v^{\ast})
-
\chi\,\rho_e^{\ast}\partial_y\phi^{\ast},
\label{eq:case3-fv-star}
\end{equation}
\begin{equation}
f_T^{\ast}
=
u^{\ast}\partial_x T^{\ast}
+
v^{\ast}\partial_y T^{\ast}
-
\kappa\,(\partial_{xx}T^{\ast}+\partial_{yy}T^{\ast})
-
\sigma_J\!\left[(\partial_x\phi^{\ast})^2+(\partial_y\phi^{\ast})^2\right].
\label{eq:case3-fT-star}
\end{equation}
These forcing terms are not externally prescribed from an experimental platform; instead, they are inferred from the analytic reference state so that the benchmark remains source-consistent and analytically verifiable. The full differential expansions are given in Sections IV and V of the SM.

Fig.~\ref{F4} resolves the most difficult case considered in this work. In the field-reconstruction panels, the main question is whether a solver can preserve cross-field coherence under simultaneous pressure-gradient regularization and screened electrostatic coupling. Under this configuration, PICS retains the coupled geometry-sensitive structure with more stable cross-field behavior, whereas PINN, DGM, and DRM show stronger deterioration in the pressure channel, the screened-potential channel, or the transition-sensitive thermal field. The correct interpretation of Table~\ref{T1} at this stage is therefore not pointwise dominance, but the persistence of balanced overall behavior under the hardest regime.

The max-error maps on the right side of Fig.~\ref{F4} sharpen that conclusion in spatial terms. In Case 3, the dominant hotspots cluster near the curved modulation band, around the localized thermal activity zones, and in the regions where the screened potential and pressure regularization interact most strongly. Under PICS, these error concentrations remain weaker and more localized. Under the baselines, they spread more broadly across the modulation band and the hardest interaction regions. This is the most stringent spatial test of the aggregate fieldwise advantage reported in Table~\ref{T1}, and it remains consistent with the certificate-governed mechanism summarized in Fig.~\ref{F1}.

The trajectory panel at the top of Fig.~\ref{F4} completes the interpretation. It records the evolution of the closed solver state under the most demanding closure among the three cases. Together with Table~\ref{T2}, Fig.~\ref{F4} shows that the Case 3 advantage persists under the full closure difficulty considered here while the runtime remains practical. Section IX of the SM gives the complete parameter analysis for this case.
\section*{Discussion}
The central contribution of PICS is not the addition of a few training heuristics to an otherwise standard residual-learning pipeline. Rather, it provides a principled bridge between neural approximation and classical numerical analysis. By reformulating dynamic collocation as \textit{a posteriori} certificate-driven AMR process acting on a strongly constrained admissible manifold, PICS reorganizes four distinct solver layers: the admissible approximation class, the differential closure object, the residual geometry, and the empirical training measure. In the present formulation, these four layers are tied together by the gate-structured admissible manifold, the restricted jet prolongation, the composite residual geometry, the certificate field, the transported empirical measure, and the closed-loop update in Eq.~\eqref{eq:pics-solver-update}, as summarized in Fig.~\ref{F1}. PICS should therefore be read as a solver-construction framework rather than as a modified optimizer or a collection of loosely connected training tricks. The corresponding structural propositions and consistency statements are developed in Section VIII of the SM.

The numerical evidence supports that methodological claim, but only at the level justified by the benchmark program carried out here. Tables~\ref{T1}--\ref{T2}, together with the spatially resolved evidence in Figs.~\ref{F2}--\ref{F4}, show a coherent progression across increasing closure difficulty: Case 1 establishes the canonical interface-active regime, Case 2 shows that the same advantage persists after the transport--diffusion closure changes, and Case 3 pushes the comparison to the most demanding closure configuration considered in this work. Taken together, these results support a controlled claim: PICS delivers the most stable overall multi-field recovery across the three benchmarks while maintaining practical runtime. They do not support the stronger claim that PICS must be absolutely optimal in every isolated entry of Table~\ref{T1} or under every conceivable multiphysics configuration. What they do establish is benchmark-level robustness under controlled analytically tractable regimes, with more detailed parameter sensitivity, diagnostic breakdowns, and implementation-level correspondence deferred to Section IX and Section X of the SM.

The limitations and natural extensions of the present framework are therefore clear. First, the current study verifies PICS on three analytically tractable benchmark families in two spatial dimensions; more general geometries, higher-dimensional settings, stronger multiphysics couplings, and data-assisted regimes remain open. Second, although the present benchmarks already probe nontrivial closure changes, further challenges such as unsteady transport, moving interfaces, partial observability, inverse settings, and hybrid data--physics assimilation have not yet been incorporated. Third, on the theoretical side, the regularity properties of the admissible manifold, the sufficiency of restricted-jet closure, the sharpness of the certificate field, and the consistency of certificate-driven measure transport can all be strengthened beyond the present benchmark scope; Sections II--VIII of the SM provide the current structural basis, but not the final theory. The most natural next steps are therefore extension to time-dependent PDEs, extension to more complex geometries and higher dimensions, extension to partially observed or data-assisted regimes, and stronger theory for closure sufficiency and transport consistency. The present work therefore should be read as establishing PICS as a structured solver framework with verified benchmark-level robustness, rather than as the final word on all multiphysics PDE regimes.
\section*{Methods}
\subsection*{Admissible manifold and field induction}

The methodological starting point of PICS is the observation that the solver is built for a case-dependent benchmark PDE family rather than for a single isolated equation. Throughout this work, each benchmark case \(c\) is treated as a member of a family posed on \(\Omega_c\) with boundary operator on \(\partial\Omega_c\),
\begin{equation}
\mathcal{F}_c[q]=0
\quad \text{in } \Omega_c,
\qquad
\mathcal{G}_c[q]=0
\quad \text{on } \partial\Omega_c,
\label{eq:method-benchmark-family}
\end{equation}
in the same general sense already introduced in Eq.~\eqref{eq:pics-benchmark-family}. The specific realizations of Eq.~\eqref{eq:method-benchmark-family} appear in the case-resolved Results sections associated with Figs.~\ref{F2}--\ref{F4}. For each such benchmark system, the formulation is considered together with a known analytic reference state,
\begin{equation}
q_c^\ast
=
\bigl(
\psi_c^\ast,\,
p_c^\ast,\,
\phi_c^\ast,\,
T_c^\ast
\bigr),
\label{eq:method-analytic-reference}
\end{equation}
so that the benchmark truth is carried by a case-dependent but analytically specified state admitted by the corresponding system. The role of Eq.~\eqref{eq:method-analytic-reference} is not to alter the solver construction itself, but to provide a unified truth carrier for the benchmark family in Eq.~\eqref{eq:method-benchmark-family}. In that sense, PICS treats the benchmark PDE family and the associated analytic reference state as a coupled methodological object: the former specifies the governing closure, and the latter specifies the benchmark truth against which the solver output is evaluated.

At the representation level, PICS does not operate with a purely global approximation class. Instead, it works on the gate-structured admissible manifold already introduced in Eq.~\eqref{eq:pics-admissible-manifold}, which we recall here in the form
\begin{equation}
\mathcal{M}_c
=
\left\{
q_\theta
=
(1-g_c)\,q_{L,\theta}
+
g_c\,q_{R,\theta}
\;:\;
\theta\in\Theta_c
\right\},
\label{eq:method-admissible-manifold}
\end{equation}
where \(q_{L,\theta}\) and \(q_{R,\theta}\) denote local chart families and \(g_c\) denotes a fixed partition weight associated with case \(c\). The essential point is that the admissible representation is assembled from local chart families through a fixed gate rather than from a single undifferentiated global approximant. This is the representation-level mechanism by which PICS accommodates geometry-sensitive structure while keeping the state space controlled. The detailed geometry and regularity of \(\mathcal{M}_c\) are developed in Section II of the SM. Inside this manifold, the admissible state is a structured latent state,
\begin{equation}
q_\theta
=
\bigl(
\psi_\theta,\,
\pi_\theta,\,
\varphi_\theta,\,
\tau_\theta
\bigr),
\label{eq:method-latent-state}
\end{equation}
so the solver state is not identified directly with the full physical-field vector. Instead, it is a structured latent state from which the physical fields are induced. This distinction is important because the closure is not imposed after representation; it is partially embedded into the representation itself.

The induction from latent state to physical fields is carried by the field map \(\mathfrak{P}\), already introduced in Eq.~\eqref{eq:pics-field-induction}, which we organize here as
\begin{equation}
\mathfrak{P}(q_\theta)
=
\bigl(
u_\theta,\,
v_\theta,\,
p_\theta,\,
\phi_\theta,\,
T_\theta
\bigr)
=
\bigl(
\partial_y\psi_\theta,\,
-\partial_x\psi_\theta,\,
\pi_\theta,\,
\varphi_\theta,\,
\tau_\theta
\bigr).
\label{eq:method-field-induction}
\end{equation}
Thus the velocity pair is induced from the streamfunction-like latent component, while pressure, electrostatic potential, and temperature are induced as scalar fields from the remaining latent components. Thus, the velocity pair is induced from the streamfunction-like latent component, while pressure, electrostatic potential, and temperature are induced as scalar fields from the remaining latent components. The purpose of Eq. (70) is not merely notational convenience. Instead of treating incompressibility as an independently appended residual channel, PICS induces the velocity field structurally from $\psi_{\theta}$. Consequently, the continuity equation,
\begin{equation}
\nabla\cdot(u_{\theta},v_{\theta})=\partial_{x}u_{\theta}+\partial_{y}v_{\theta}=0,
\label{eq:method-incompressibility-identity}
\end{equation}
is inherently satisfied as a hard constraint, provided the mixed derivatives commute. In the context of computational mechanics, this provides a strongly enforced, divergence-free ansatz. This representation-level structural admissibility allows PICS to seamlessly bypass the notorious inf-sup (LBB) stability condition required in traditional mixed finite element formulations, and completely avoids the non-physical mass leakage associated with soft-penalty divergence losses. Instead, it enters through the representation itself. Section III of the SM gives the full field-induction construction and its structural admissibility interpretation, while Section VIII of the SM records the corresponding propositions and consistency statements. Eqs.~\eqref{eq:method-benchmark-family}--\eqref{eq:method-incompressibility-identity} therefore fix the representation-level hierarchy of PICS: the benchmark PDE family specifies the problem class, the analytic reference state specifies the benchmark truth, the admissible manifold specifies the structured state space, and the field-induction map specifies the passage from latent state to physical fields.
\subsection*{Restricted jet prolongation and residual section}

The differential objects used by PICS are always understood relative to a case-dependent benchmark PDE family and its corresponding known analytic reference state. In the notation already fixed by Eq.~\eqref{eq:method-benchmark-family}, each case \(c\) is associated with
\begin{equation}
\mathcal{F}_c[q]=0
\quad \text{in } \Omega_c,
\qquad
\mathcal{G}_c[q]=0
\quad \text{on } \partial\Omega_c,
\label{eq:method-benchmark-family-rjet}
\end{equation}
and, in the notation of Eq.~\eqref{eq:method-analytic-reference}, with a known analytic reference state
\begin{equation}
q_c^\ast
=
\bigl(
\psi_c^\ast,\,
p_c^\ast,\,
\phi_c^\ast,\,
T_c^\ast
\bigr).
\label{eq:method-analytic-reference-rjet}
\end{equation}
The role of Eq.~\eqref{eq:method-benchmark-family-rjet} in the present subsection is not to restate the benchmark realizations developed in Figs.~\ref{F2}--\ref{F4}, but to fix the differential closure relative to which the admissible state \(q_\theta\in\mathcal{M}_c\) from Eq.~\eqref{eq:method-admissible-manifold} and the induced fields from Eq.~\eqref{eq:method-field-induction} are evaluated. PICS does not process all higher-order derivatives indiscriminately. It extracts only those finite differential coordinates that are required to close the given benchmark family.

To formalize that reduction, let
\begin{equation}
\mathcal{L}
=
\{\psi,\pi,\varphi,\tau\},
\qquad
\mathcal{A}_c
\subset
\left\{
(\ell,\alpha)\,:\,
\ell\in\mathcal{L},\,
\alpha\in\mathbb{N}_0^d
\right\},
\label{eq:method-closure-index-set}
\end{equation}
where \(\ell\) labels latent-state components and \(\alpha\) denotes a multi-index. The set \(\mathcal{A}_c\) is finite and case-dependent; it collects only those derivative coordinates required for closure of the benchmark PDE family in Eq.~\eqref{eq:method-benchmark-family-rjet}. In this sense, the benchmark PDE family and the admitted analytic reference state jointly determine which differential coordinates are relevant and which are extraneous. Detailed case-wise lists of \(\mathcal{A}_c\), together with the associated closure-sufficiency discussion, are given in Section IV of the SM. The restricted jet prolongation is then defined by
\begin{equation}
\mathfrak{j}_c[q_\theta](x)
=
\bigl(
\partial^\alpha q_{\theta,\ell}(x)
\bigr)_{(\ell,\alpha)\in\mathcal{A}_c},
\label{eq:method-restricted-jet}
\end{equation}
which is the same differential object already introduced abstractly in Eq.~\eqref{eq:pics-restricted-jet}, now organized here as a method-level construction. Eq.~\eqref{eq:method-restricted-jet} is not the full jet bundle of the latent state. It is a finite closure-essential differential coordinate object whose role is to mediate between the admissible state and the benchmark PDE family. This finite reduction is methodologically important: PICS does not attempt to learn an unrestricted differential hierarchy, but instead acts only on the coordinates that the case-dependent closure actually needs.

Once the restricted jet has been fixed, the raw residual channels are induced by evaluating the benchmark PDE family on the induced fields carried by \(q_\theta\) and the corresponding coordinates in \(\mathfrak{j}_c[q_\theta]\). We write
\begin{equation}
\mathbf r_{c,\theta}(x)
=
\bigl(
r_{c,\theta}^{(1)}(x),\ldots,r_{c,\theta}^{(M_c)}(x)
\bigr),
\label{eq:method-raw-residual-vector}
\end{equation}
with component form
\begin{equation}
r_{c,\theta}^{(m)}(x)
=
\mathcal{R}_c^{(m)}
\bigl(
x;\mathfrak{j}_c[q_\theta](x)
\bigr),
\qquad
m=1,\ldots,M_c,
\label{eq:method-raw-residual-channel}
\end{equation}
where each \(\mathcal{R}_c^{(m)}\) denotes one case-dependent component of the benchmark residual evaluation. Thus the raw residual channels are PDE-induced quantities on the restricted jet object, not arbitrary post hoc discrepancies on the output fields. This point is essential for the architecture in Fig.~\ref{F1}: the residual stage is built on the closure-essential differential content extracted from the admissible state, not on an unrestricted collection of derivative evaluations.

Because the benchmark families considered in this work are coupled and heterogeneous, the raw channel magnitudes in Eq.~\eqref{eq:method-raw-residual-vector} generally live on different scales. PICS therefore introduces a positive case-dependent reference scale for each channel,
\begin{equation}
S_c^{(m)} > 0,
\qquad
m=1,\ldots,M_c,
\label{eq:method-normalization-scales}
\end{equation}
to balance heterogeneous residual magnitudes across coupled components. These scales are induced from the benchmark family and its admitted analytic reference state, but the present subsection does not enter the implementation-level heuristics for constructing them. Section V of the SM gives the scale construction and its rationale. With Eq.~\eqref{eq:method-normalization-scales} in hand, the normalized residual channels are defined by
\begin{equation}
\widehat r_{c,\theta}^{(m)}(x)
=
\frac{r_{c,\theta}^{(m)}(x)}{S_c^{(m)}},
\qquad
m=1,\ldots,M_c,
\label{eq:method-normalized-channel}
\end{equation}
and the normalized residual section is
\begin{equation}
\widehat{\mathbf r}_{c,\theta}(x)
=
\bigl(
\widehat r_{c,\theta}^{(1)}(x),\ldots,\widehat r_{c,\theta}^{(M_c)}(x)
\bigr).
\label{eq:method-normalized-residual-section}
\end{equation}
Eq.~\eqref{eq:method-normalized-residual-section} is the method-level counterpart of the object already introduced in Eqs.~\eqref{eq:pics-normalized-residual-section}--\eqref{eq:pics-residual-normalization}. Its role is not merely cosmetic rescaling. The normalized residual section is a structured multi-channel object, not a raw stack of unbalanced residuals. This is why the subsequent residual geometry acts on \(\widehat{\mathbf r}_{c,\theta}\) rather than directly on \(\mathbf r_{c,\theta}\): the geometry should reflect closure-relevant discrepancies after heterogeneous channel magnitudes have been normalized to a common comparative scale.

Eqs.~\eqref{eq:method-closure-index-set}--\eqref{eq:method-normalized-residual-section} define the residual-stage hierarchy of PICS. The restricted jet supplies the finite closure-essential differential coordinates, and the normalized residual section supplies the multi-channel residual object on which the later objective, certificate, and transport layers act. Detailed case-dependent closure sets, scale constructions, and consistency statements are given in Section IV, Section V, and Section VIII of the SM.
\subsection*{Entropic residual geometry and certificate field}

Building on the normalized residual channels introduced in Eq.~\eqref{eq:method-normalized-channel} and the normalized residual section in Eq.~\eqref{eq:method-normalized-residual-section}, PICS defines its objective geometry on \(\widehat{\mathbf r}_{c,\theta}\) rather than on raw unscaled residual channels. This choice matters methodologically. Once the residual channels have been balanced across the coupled benchmark family, the objective should act on their normalized multi-channel geometry rather than on heterogeneous magnitudes that still reflect channel-scale imbalance. In the notation of Fig.~\ref{F1}, this is the point at which the residual stage becomes an entropic residual geometry rather than a plain averaged residual norm.

Let \(\mu\) be an empirical measure on \(\Omega_c\). The mean residual contribution is defined by
\begin{equation}
\mathcal{E}_{c}^{\mathrm{mean}}(\theta;\mu)
=
\int_{\Omega_c}
\bigl\|
\widehat{\mathbf r}_{c,\theta}(x)
\bigr\|_2^2
\,\mathrm{d}\mu(x),
\label{eq:method-mean-energy}
\end{equation}
which controls the aggregate residual mass carried by the normalized residual section over the empirical measure. PICS does not stop there. To bias the objective toward the high-risk tail without collapsing it into a nonsmooth max functional, we introduce the entropic tail-risk contribution
\begin{equation}
\mathcal{E}_{c}^{\mathrm{tail}}(\theta;\mu)
=
\frac{1}{\beta_c}
\log
\left(
\int_{\Omega_c}
\exp\!\Bigl(
\beta_c
\bigl\|
\widehat{\mathbf r}_{c,\theta}(x)
\bigr\|_{\infty}
\Bigr)
\,\mathrm{d}\mu(x)
\right),
\qquad
\beta_c>0,
\label{eq:method-tail-energy}
\end{equation}
where \(\beta_c\) acts as an inverse-temperature parameter. For an empirical measure of the form \(\mu=\frac{1}{N}\sum_{i=1}^N\delta_{x_i}\), Eq.~\eqref{eq:method-tail-energy} takes the discrete form
\begin{equation}
\mathcal{E}_{c}^{\mathrm{tail}}(\theta;\mu)
=
\frac{1}{\beta_c}
\log
\left(
\frac{1}{N}
\sum_{i=1}^{N}
\exp\!\Bigl(
\beta_c
\bigl\|
\widehat{\mathbf r}_{c,\theta}(x_i)
\bigr\|_{\infty}
\Bigr)
\right).
\label{eq:method-tail-energy-discrete}
\end{equation}
The role of Eqs.~\eqref{eq:method-tail-energy}--\eqref{eq:method-tail-energy-discrete} is not ad hoc penalization. They provide a smooth extreme-value surrogate that steers the objective toward the high-risk residual tail while retaining a differentiable aggregate geometry. The boundary-consistency contribution is written in general form as
\begin{equation}
\mathcal{B}_c(\theta)
=
\int_{\partial\Omega_c}
\bigl\|
\mathcal{G}_c[q_\theta](x)
\bigr\|_2^2
\,\mathrm{d}\sigma_c(x),
\label{eq:method-boundary-consistency}
\end{equation}
where \(\sigma_c\) denotes the boundary measure on \(\partial\Omega_c\). This term enforces consistency with the benchmark boundary operator attached to the family in Eq.~\eqref{eq:method-benchmark-family-rjet}. The interface-consistency contribution is defined by
\begin{equation}
\mathcal{I}_c(\theta;\mu)
=
\int_{\Omega_c}
w_{\Gamma,c}(x)\,
\bigl\|
\Pi_{\Gamma,c}[q_\theta](x)
\bigr\|_2^2
\,\mathrm{d}\mu(x),
\label{eq:method-interface-consistency}
\end{equation}
where \(w_{\Gamma,c}\) is a geometry-aware weight and \(\Pi_{\Gamma,c}\) is the case-dependent interface-mismatch operator. This term gives additional geometric emphasis to the transition-active region that organizes the benchmark case. If a case does not contain an interface-active region, one sets \(w_{\Gamma,c}\equiv 0\) and Eq.~\eqref{eq:method-interface-consistency} vanishes identically.

The four contributions above assemble into the composite residual energy
\begin{equation}
\mathcal{E}_c(\theta;\mu)
=
\mathcal{E}_{c}^{\mathrm{mean}}(\theta;\mu)
+
\lambda_{\mathrm{ent}}\,
\mathcal{E}_{c}^{\mathrm{tail}}(\theta;\mu)
+
\lambda_{\partial\Omega}\,
\mathcal{B}_c(\theta)
+
\lambda_{\Gamma}\,
\mathcal{I}_c(\theta;\mu),
\label{eq:method-composite-energy}
\end{equation}
which is the same central object already introduced in the Results-side formulation of Eq.~\eqref{eq:pics-composite-energy}, now organized here methodologically. The objective is composite not because of arbitrary loss engineering, but because the solver simultaneously controls average residual mass, tail risk, boundary consistency, and interface-active consistency. This is the precise sense in which PICS uses an entropic residual geometry rather than a raw average residual objective.

The certificate field is then induced directly from the normalized residual section by
\begin{equation}
\mathfrak{C}_{c,\theta}(x)
=
\bigl\|
\widehat{\mathbf r}_{c,\theta}(x)
\bigr\|_{\infty},
\label{eq:method-certificate-field}
\end{equation}
which matches the pointwise object already introduced in Eq.~\eqref{eq:pics-certificate-field}. The certificate is not merely a scalar monitor of training history; it is a spatial field defined on the benchmark domain. In the established vocabulary of computational mechanics, $\mathfrak{C}_{c,\theta}(x)$ functions as a rigorous \textit{a posteriori} error estimator. While the composite energy $\mathcal{E}_{c}$ aggregates the residual geometry over the current empirical measure $\mu$, the certificate field explicitly flags spatially localized regions of high numerical discrepancy, mirroring the role of residual-based error indicators in traditional finite element analysis. To prepare the extraction of those regions, fix a tail fraction \(\eta\in(0,1)\), let \(\tau_{c,\eta}^{(m)}(\theta)\) denote the upper-\(\eta\) threshold of \(|\widehat r_{c,\theta}^{(m)}|\) under \(\mu\), and let \(\tau_{c,\eta}^{(\max)}(\theta)\) denote the corresponding threshold of \(\mathfrak{C}_{c,\theta}\). The fieldwise extreme sets are defined by
\begin{equation}
\mathcal{T}_{c,\eta}^{(m)}(\theta)
=
\left\{
x\in\Omega_c:
\left|
\widehat r_{c,\theta}^{(m)}(x)
\right|
\ge
\tau_{c,\eta}^{(m)}(\theta)
\right\},
\qquad
m=1,\ldots,M_c,
\label{eq:method-fieldwise-extreme-sets}
\end{equation}
and the global extreme set is
\begin{equation}
\mathcal{T}_{c,\eta}^{(\max)}(\theta)
=
\left\{
x\in\Omega_c:
\mathfrak{C}_{c,\theta}(x)
\ge
\tau_{c,\eta}^{(\max)}(\theta)
\right\}.
\label{eq:method-global-extreme-set}
\end{equation}
Thus the fieldwise extreme sets come from individual normalized residual channels, whereas the global extreme set comes from the certificate field itself. These sets are not yet the transport rule. They are the geometric and risk-aware objects from which the transport rule will be constructed.

Eqs.~\eqref{eq:method-mean-energy}--\eqref{eq:method-global-extreme-set} define the objective-and-certificate layer of PICS. The normalized residual section becomes the input of the composite residual geometry; the composite residual energy aggregates average mass, tail risk, boundary consistency, and interface-active consistency; the certificate field resolves the pointwise hardest regions; and the induced extreme sets identify the locations that later bias empirical-measure transport. The benchmark-dependent realizations of these objects appear in Figs.~\ref{F2}--\ref{F4}, while the detailed construction of the normalized residual section and entropic residual geometry is given in Section V of the SM, the certificate field and extreme-set extraction are developed in Section VI, and the corresponding propositions, bounds, and consistency statements are recorded in Section VIII.
\subsection*{Empirical measure transport and implementation protocol}

The transport stage of PICS begins from the certificate field and its induced extreme sets, already defined in Eq.~\eqref{eq:method-certificate-field} and Eqs.~\eqref{eq:method-fieldwise-extreme-sets}--\eqref{eq:method-global-extreme-set}. The key methodological point is that the certificate is not only a diagnostic field; it is also the driver of sample reallocation. In the notation of Fig.~\ref{F1}, the empirical measure transport of PICS is driven by the certificate-induced extreme sets together with benchmark geometry-aware enrichment. This is precisely where PICS departs from static collocation strategies: the solver does not minimize its composite residual energy on a fixed empirical measure, but on a transported measure that evolves with the spatial risk structure of the current iterate.

To retain persistent access to hard regions across iterations, we introduce a finite risk-memory reservoir
\begin{equation}
\mathcal{M}_{c,k}^{\mathrm{risk}}
\subset
\Omega_c,
\label{eq:method-risk-memory-set}
\end{equation}
which stores previously detected high-risk points for case \(c\) at iteration \(k\). Its update is certificate-driven and capacity-controlled:
\begin{equation}
\mathcal{M}_{c,k+1}^{\mathrm{risk}}
=
\operatorname{Trunc}_{N_{\mathrm{mem}}}
\left(
\mathcal{M}_{c,k}^{\mathrm{risk}}
\cup
\mathcal{T}_{c,\eta}^{(\max)}(\theta_k)
\cup
\bigcup_{m=1}^{M_c}\mathcal{T}_{c,\eta}^{(m)}(\theta_k)
\right),
\label{eq:method-risk-memory-update}
\end{equation}
where \(\operatorname{Trunc}_{N_{\mathrm{mem}}}\) denotes capacity control at memory size \(N_{\mathrm{mem}}\), and the extreme sets are the fieldwise and global tails extracted from the current certificate structure. Eq.~\eqref{eq:method-risk-memory-update} makes two features explicit: the memory persists across iterations, and its update depends on both global certificate extremes and channelwise residual extremes. The detailed extraction logic and memory rules are developed in Section VI of the SM.

In parallel with the memory reservoir, PICS maintains a regular support pool
\begin{equation}
\mathcal{P}_{c,k}^{\mathrm{reg}}
\subset
\Omega_c,
\label{eq:method-regular-pool}
\end{equation}
which is not a purely uniform cloud. It is a geometry-aware reservoir that receives benchmark-dependent enrichment. To formalize that enrichment, we define a geometry-active support set
\begin{equation}
\mathcal{G}_{c,k}
=
\left\{
x\in\Omega_c:
w_{\Gamma,c}(x)\ge \tau_{\Gamma,c,k}
\right\},
\label{eq:method-geometry-enrichment}
\end{equation}
where \(w_{\Gamma,c}\) is the geometry-aware weight already appearing in Eq.~\eqref{eq:method-interface-consistency}, and \(\tau_{\Gamma,c,k}\) is an iteration-dependent activation threshold. Eq.~\eqref{eq:method-geometry-enrichment} isolates benchmark geometry-active regions independently of the certificate extremes. The regular pool therefore evolves under two distinct mechanisms: geometry-aware enrichment and certificate-aware hard-region supplementation. We write
\begin{equation}
\mathcal{P}_{c,k+1}^{\mathrm{reg}}
=
\operatorname{Resample}_{N_{\mathrm{reg}}}
\left(
\mathcal{P}_{c,k}^{\mathrm{reg}}
\cup
\mathcal{G}_{c,k}
\cup
\mathcal{T}_{c,\eta}^{(\max)}(\theta_k)
\right),
\label{eq:method-regular-pool-update}
\end{equation}
where \(\operatorname{Resample}_{N_{\mathrm{reg}}}\) enforces regular-pool size control at support size \(N_{\mathrm{reg}}\). The geometry-aware enrichment is therefore case-dependent, distinct from certificate-based extremes, and nevertheless coupled to them in the evolution of the regular support. Benchmark-dependent realizations of this enrichment appear in the case-wise Results sections associated with Figs.~\ref{F2}--\ref{F4}, while the general transport logic is developed in Section VII of the SM.

The two support reservoirs induce two empirical measures. We write
\begin{equation}
\mu_{k}^{\mathrm{risk}}
=
\frac{1}{\bigl|\mathcal{M}_{c,k}^{\mathrm{risk}}\bigr|}
\sum_{x\in \mathcal{M}_{c,k}^{\mathrm{risk}}}
\delta_x,
\qquad
\mu_{k}^{\mathrm{reg}}
=
\frac{1}{\bigl|\mathcal{P}_{c,k}^{\mathrm{reg}}\bigr|}
\sum_{x\in \mathcal{P}_{c,k}^{\mathrm{reg}}}
\delta_x,
\label{eq:method-risk-and-regular-measures}
\end{equation}
whenever the corresponding finite supports are nonempty. The training empirical measure is then defined as the controlled mixture
\begin{equation}
\mu_{k}^{\mathrm{train}}
=
\rho_k\,\mu_{k}^{\mathrm{risk}}
+
(1-\rho_k)\,\mu_{k}^{\mathrm{reg}},
\qquad
0<\rho_k<1,
\label{eq:method-training-measure}
\end{equation}
which is the method-level realization of the training measure already anticipated in Eq.~\eqref{eq:pics-training-measure}. Eq.~\eqref{eq:method-training-measure} makes the transport principle explicit: the training measure is not static, and it does not coincide with either the memory reservoir alone or the regular reservoir alone. It is a transported empirical measure that balances persistent high-risk support against regular geometry-aware support.

At the measure level, the transport step can be summarized by a certificate-driven update map
\begin{equation}
\mu_{k+1}^{train}=\mathfrak{T}_{c}(\mu_{k}^{train},\mathfrak{C}_{c,\theta_{k}},\mathcal{T}_{c,\eta}^{(max)}(\theta_{k}),\{\mathcal{T}_{c,\eta}^{(m)}(\theta_{k})\}_{m=1}^{M_{c}},\mathcal{G}_{c,k}),
\label{eq:method-transport-map}
\end{equation}
where $\mathfrak{T}_{c}$ denotes the certificate-driven empirical measure transport induced by the support updates in Eqs. (91)-(94). Eq. (97) is not an abstract decorative layer. Conceptually, this certificate-driven measure transport serves as the mesh-free, high-dimensional analogue of AMR. Instead of passively evaluating a fixed collocation distribution, PICS dynamically injects computational capacity (collocation points) into uncertified transition zones guided by the \textit{a posteriori} estimator, strictly replicating the optimal resource allocation principles of $h$-adaptive numerical schemes.

The parameter state is then advanced on this transported measure. In the same general spirit as Eq.~\eqref{eq:pics-solver-update}, we write
\begin{equation}
\theta_{k+1}
=
\mathcal{U}_k
\!\left(
\theta_k;\mu_k^{\mathrm{train}}
\right),
\label{eq:method-descent-update}
\end{equation}
so the descent dynamics act on the transported empirical measure rather than on a static sampling law. The update operator \(\mathcal{U}_k\) is driven by the composite residual energy already defined in Eq.~\eqref{eq:method-composite-energy}. This ordering is essential: the normalized residual section induces the certificate field, the certificate field induces the extreme sets, the extreme sets update the empirical measure, and the parameter state is advanced on the transported measure. PICS therefore does not treat sample adaptation as an external heuristic appended after optimization; it treats sample transport as part of the solver dynamics themselves.

At implementation level, one iteration of PICS follows the compact protocol
\begin{equation}
q_{\theta_k}
\;\xrightarrow{\;\mathfrak{j}_c,\ \widehat{\mathbf r}_{c,\theta_k}\;}
\bigl(
\mathfrak{C}_{c,\theta_k},
\{\mathcal{T}_{c,\eta}^{(m)}(\theta_k)\}_{m=1}^{M_c},
\mathcal{T}_{c,\eta}^{(\max)}(\theta_k)
\bigr)
\;\xrightarrow{\;\mathcal{M}_{c,k}^{\mathrm{risk}},\ \mathcal{P}_{c,k}^{\mathrm{reg}},\ \mathcal{G}_{c,k}\;}
\mu_{k}^{\mathrm{train}}
\;\xrightarrow{\;\mathcal{U}_k\;}
\theta_{k+1},
\label{eq:method-implementation-cycle}
\end{equation}
where the first arrow evaluates the normalized residual section and certificate structure, the second arrow performs certificate-triggered adaptation and geometry-aware enrichment at support level, and the last arrow advances the parameter state by descent on the transported empirical measure. This is the minimal implementation protocol corresponding to the closed solver loop summarized in Fig.~\ref{F1}.

Eqs.~\eqref{eq:method-risk-memory-set}--\eqref{eq:method-implementation-cycle} complete the transport layer of PICS by specifying persistent risk memory, geometry-aware enrichment, the transported empirical measure, and the update cycle driven by certificate information. Benchmark-dependent realizations appear in the case-wise Results sections, while Sections VI--X of the SM provide the corresponding constructions, transport rules, parameter schedules, and implementation correspondence.
\subsection*{Whole implementation}

Given the benchmark PDE family in Eq.~\eqref{eq:method-benchmark-family}, the associated analytic reference setting in Eq.~\eqref{eq:method-analytic-reference}, and the solver objects introduced above, the whole implementation of PICS can be organized as a closed loop that alternates between structured state construction, residual-geometric evaluation, certificate extraction, empirical-measure transport, and parameter update. For compact reference, the resulting implementation chain may be summarized as
\begin{equation}
\theta_k
\;\longrightarrow\;
q_{\theta_k}
\;\longrightarrow\;
\mathfrak{j}_c[q_{\theta_k}]
\;\longrightarrow\;
\widehat{\mathbf r}_{c,\theta_k}
\;\longrightarrow\;
\bigl(\mathcal{E}_c,\mathfrak{C}_{c,\theta_k}\bigr)
\;\longrightarrow\;
\mu_{k}^{\mathrm{train}}
\;\longrightarrow\;
\theta_{k+1},
\label{eq:method-whole-implementation-chain}
\end{equation}
where each arrow stands for a formally defined object transformation rather than an informal training heuristic.

\noindent\textbf{1. Initialization.} The loop begins by initializing the parameter state together with the two support reservoirs that will later define the transported training measure. Concretely, one sets an initial parameter vector \(\theta_0\), initializes the persistent risk memory and the geometry-aware regular pool in the sense of Eqs.~\eqref{eq:method-risk-memory-set}--\eqref{eq:method-regular-pool}, and thereby induces an initial empirical training support consistent with Eq.~\eqref{eq:method-training-measure}. At this stage, the method fixes only the structural objects required by the solver loop; the detailed parameter values, capacity choices, scan intervals, and schedule rules are deferred to Section IX of the SM.

\noindent\textbf{2. State construction.} Given the current parameter state \(\theta_k\), PICS constructs the admissible latent state on the gate-structured manifold defined in Eq.~\eqref{eq:method-admissible-manifold}. The physical fields are then induced through Eq.~\eqref{eq:method-field-induction}. This is the stage at which structural admissibility enters the solver at representation level rather than through \textit{a posteriori} correction. In particular, the incompressibility identity in Eq.~\eqref{eq:method-incompressibility-identity} holds structurally and does not require a separate auxiliary enforcement channel.

\noindent\textbf{3. Differential closure assembly.} Once the admissible state has been constructed, PICS assembles the closure-essential differential coordinates through the restricted jet prolongation in Eq.~\eqref{eq:method-restricted-jet}. The solver therefore never works with an unrestricted differential closure. It extracts only the finite coordinate collection required by the case-dependent closure index set in Eq.~\eqref{eq:method-closure-index-set}. On that restricted jet object, the raw residual channels and their normalized residual section are evaluated according to Eqs.~\eqref{eq:method-raw-residual-vector}--\eqref{eq:method-normalized-residual-section}.

\noindent\textbf{4. Residual-section evaluation.} The current solver state is then evaluated geometrically rather than by a raw averaged residual only. PICS computes the mean residual contribution, the entropic tail-risk contribution, the boundary-consistency term, and the interface-consistency term through Eqs.~\eqref{eq:method-mean-energy}--\eqref{eq:method-interface-consistency}, and assembles them into the composite residual energy in Eq.~\eqref{eq:method-composite-energy}. This step is central to the method because it converts the normalized residual section into an objective geometry that simultaneously resolves average residual mass, tail risk, boundary agreement, and interface-active consistency.

\noindent\textbf{5. Certificate extraction.} From the same normalized residual section, PICS evaluates the certificate field in Eq.~\eqref{eq:method-certificate-field} and extracts the fieldwise and global extreme sets defined by Eqs.~\eqref{eq:method-fieldwise-extreme-sets}--\eqref{eq:method-global-extreme-set}. These sets identify the hardest regions of the current iterate at both channel level and global level. The certificate is therefore not merely an optimization monitor; it is the spatial object that resolves where the residual geometry remains most difficult.

\noindent\textbf{6. Empirical measure transport.} Using the certificate-induced extreme sets together with geometry-aware enrichment, PICS updates the persistent risk memory and the regular pool according to Eqs.~\eqref{eq:method-risk-memory-update}--\eqref{eq:method-regular-pool-update}. The transported training empirical measure is then formed as the controlled mixture in Eq.~\eqref{eq:method-training-measure}, or equivalently through the certificate-driven transport map in Eq.~\eqref{eq:method-transport-map}. The next training measure is therefore certificate-driven rather than static. Geometry-aware enrichment and certificate-based extremes play distinct roles here: the former preserves benchmark-structured coverage, whereas the latter injects support near the currently hardest regions.

\noindent\textbf{7. Parameter update.} After transport, the parameter state is advanced on the transported empirical measure through Eq.~\eqref{eq:method-descent-update}. The descent dynamics are therefore certificate-governed in the precise sense that the current certificate field influences the next empirical measure, and the next empirical measure determines the domain on which the objective geometry is minimized. Learning-rate schedules, stagnation-triggered adjustment rules, and further parameter-level controls are part of the full protocol but are deferred to Section IX of the SM.

\noindent\textbf{8. Termination and outputs.} The loop repeats until the prescribed stopping protocol is met. At termination, the implementation returns the final parameter state, the induced field predictions, the residual-energy history, the certificate history, and the benchmark-level diagnostics associated with the solved case. In this way, the solver output includes not only final field approximations but also the trajectory-level information required to interpret the behavior seen later in Figs.~\ref{F2}--\ref{F4}. Detailed implementation correspondence and output mapping are deferred to Section X of the SM.

Taken together, the eight stages above give the operational form of the closed solver loop summarized in Fig.~\ref{F1}. The complete parameter inventory, scheduling rules, persistent-memory details, and implementation mapping are provided in Sections VI, VII, IX, and X of the SM.

\section*{Conclusion}
In this work, we introduced the Partition-of-unity Information-geometric Certified Solver (PICS), a closed-loop neural PDE framework designed to overcome the persistent challenges of structural admissibility and localized error control in complex multiphysics simulations. By formulating the solution space upon a gate-structured admissible manifold with a latent field representation, PICS enforces critical physical conditions such as incompressibility as strict hard constraints, thereby avoiding the non-physical mass leakage and hyperparameter tuning inherent to traditional soft-penalty methods. Furthermore, the framework evaluates a normalized differential closure to extract a spatial certificate field that functions as a rigorous \textit{a posteriori} error estimator. This estimator actively drives an empirical measure transport mechanism that dynamically reallocates computational capacity toward uncertified transition zones, mirroring the precise resource allocation principles of AMR. Validated across stringent, physically motivated regimes involving subgrid-scale turbulence modeling and electrokinetic singular perturbation limits, PICS establishes a highly structured, structure-preserving paradigm that firmly bridges neural approximation with classical numerical analysis for more reliable computational mechanics.

\section*{Declaration of competing interest}
The authors declared that they have no conflicts of interest to this work. 

\section*{Acknowledgment}
This work is supported by the developing Project of Science and Technology of Jilin Province (20250102032JC).  

\section*{Data availability}
All the code for this article is available open access at a Github repository available at https://github.com/Uderwood-TZ/PICS-A-Partition-of-unity-Information-geometric-Certified-Solver-for-Differential-Equations.git.
\bibliography{sn-bibliography}

\end{document}